\documentclass[10pt,onecolumn,letterpaper]{article}

\usepackage{cvpr}
\usepackage{float}
\usepackage{times}
\usepackage{epsfig}
\usepackage{graphicx}
\usepackage{amsmath}
\usepackage{amssymb}
\usepackage{commath}
\usepackage{siunitx}
\usepackage[justification=centering]{subcaption}

\usepackage[margin=1.5in]{geometry}
\usepackage[pagebackref=true,breaklinks=true,letterpaper=true,colorlinks,bookmarks=false]{hyperref}

\graphicspath{ {./images/} }

\usepackage[breaklinks=true,bookmarks=false]{hyperref}

\cvprfinalcopy 

\def\cvprPaperID{****} 

\begin{document}

\title{Microscopy Image Restoration using Deep Learning on W2S}

\author{Martin Chatton$^*$}

\maketitle

\begin{abstract}
We leverage deep learning techniques to jointly denoise and super-resolve biomedical images acquired with fluorescence microscopy. We develop a deep learning algorithm based on the networks and method described in the recent W2S paper to solve a joint denoising and super-resolution problem. Specifically, we address the restoration of SIM images from widefield images. Our TensorFlow model is trained on the W2S dataset of cell images and is made accessible online in this repository: \href{https://github.com/mchatton/w2s-tensorflow}{https://github.com/mchatton/w2s-tensorflow}. On test images, the model shows a visually-convincing denoising and increases the resolution by a factor of two compared to the input image. For a 512 $\times$ 512 image, the inference takes less than 1 second on a Titan X GPU and about 15 seconds on a common CPU. We further present the results of different variations of losses used in training.
\end{abstract}

\let\thefootnote\relax\footnotetext{$^*$Supervised by Ruofan Zhou and Majed El Helou at IVRL, EPFL, Switzerland.}

\section{Introduction}

In this project, we focus on image restoration for biomedical microscopy, and in particular on super-resolution fluorescence microscopy. One significant advancement in that field is the invention of Structured-Illumination Microscopy (SIM) \cite{SIM} that enabled doubling of the resolution of acquisitions by exposing the sample multiple times to a structured pattern that induces a Moiré effect that makes details of higher spatial frequency appear on the acquisition. Shifting and rotating these patterns and then combining the images allows to double the resolution (and the largest frequency content), compared to a single acquisition. However, this enhancement comes at a cost. Indeed the sample must be exposed multiple times, which can induce photobleaching and damage living cells \cite{phototoxicity}. In addition,  the patterns need to be changed between each exposure, which is time-consuming and incompatible with living specimens displaying spatial movements, and incompatible with video acquisition.

This project aims to recreate the enhancements obtained with SIM using deep learning. This would allow to obtain the same image enhancements but using only a single microscopy image acquisition, which would better preserve the observed cells and be more time-efficient.

Obtaining similar results as SIM using image processing would require both efficient upsampling and denoising. Denoising images is a problem that is well addressed by deep leaning. Multiple methods have been developed on the use of Convolutional Neural-Networks (CNNs) for image denoising, mainly started by Zhang \etal \cite{residualdenoising} who introduced the use of residual learning. Further research introduced the use of recursive units to represent persistent memory \cite{memnet} and very recent approaches use feature attention learning \cite{featureattention} and Bayesian priors \cite{blinddenoising}.

Image super-resolution, also called Single Image Super-Resolution (SISR), where one single low resolution image is transformed into a higher resolution version, was first introduced using CNNs by Dong \etal \cite{SR-CNN}. Most recent architectures \cite{enhancenet, SRGAN, ESRGAN,residual-dense} make use of residual learning that has been proven to be effective for both super-resolution and denoising individually.

Even if both problems share similarities in both their aim and solution techniques, they are not effective when the problem is joint as in our case, as deep learning algorithms for super-resolution are proven to be very sensitive to noise \cite{sr-attack}. One could apply a denoising algorithm followed by a super-resolution one, but it would be inefficient. The reason super-resolution algorithms are not able to handle noise is because they are not trained on datasets of noisy images. The problem of joint denoising and super-resolution can hence only be effectively addressed by using an appropriate dataset.

For that purpose, we use the \textit{Widefield to SIM} (W2S) \cite{W2S} dataset which consists of grayscale image acquisitions of human cell samples with different fields of view and acquisition wavelengths, before any processing and after the SIM process. This will allow the network to automatically recover a high-definition and denoised version of the image as if it were acquired with SIM. In this dataset, the images size is 512 $\times$ 512 pixels in low resolution and 1024 $\times$ 1024 in high resolution. The SIM process produced images of twice the resolution of the first image.

\begin{figure}[h]
\begin{minipage}[t]{0.49\linewidth}
\centering
\includegraphics[width=\linewidth]{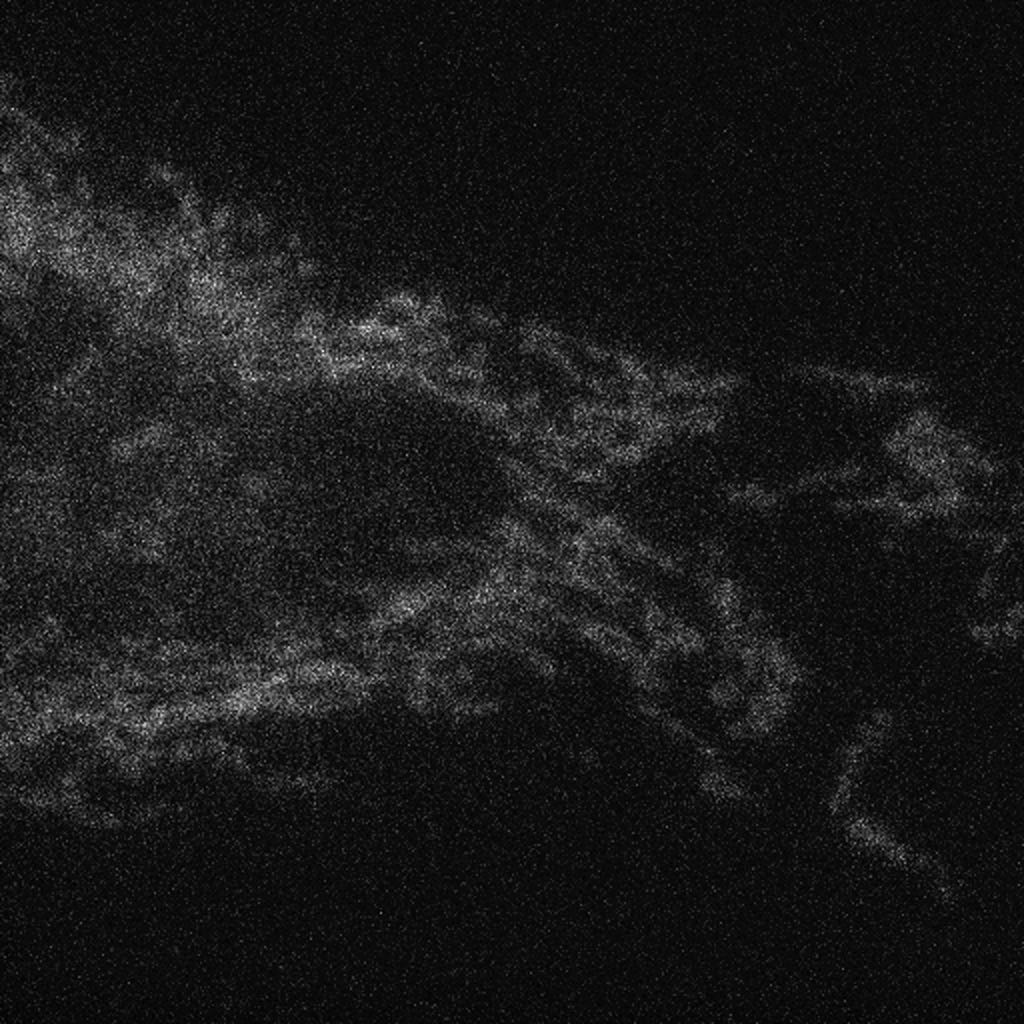}

Low resolution (input)
\end{minipage}%
\hfill
\begin{minipage}[t]{0.49\linewidth}
\centering
\includegraphics[width=\linewidth]{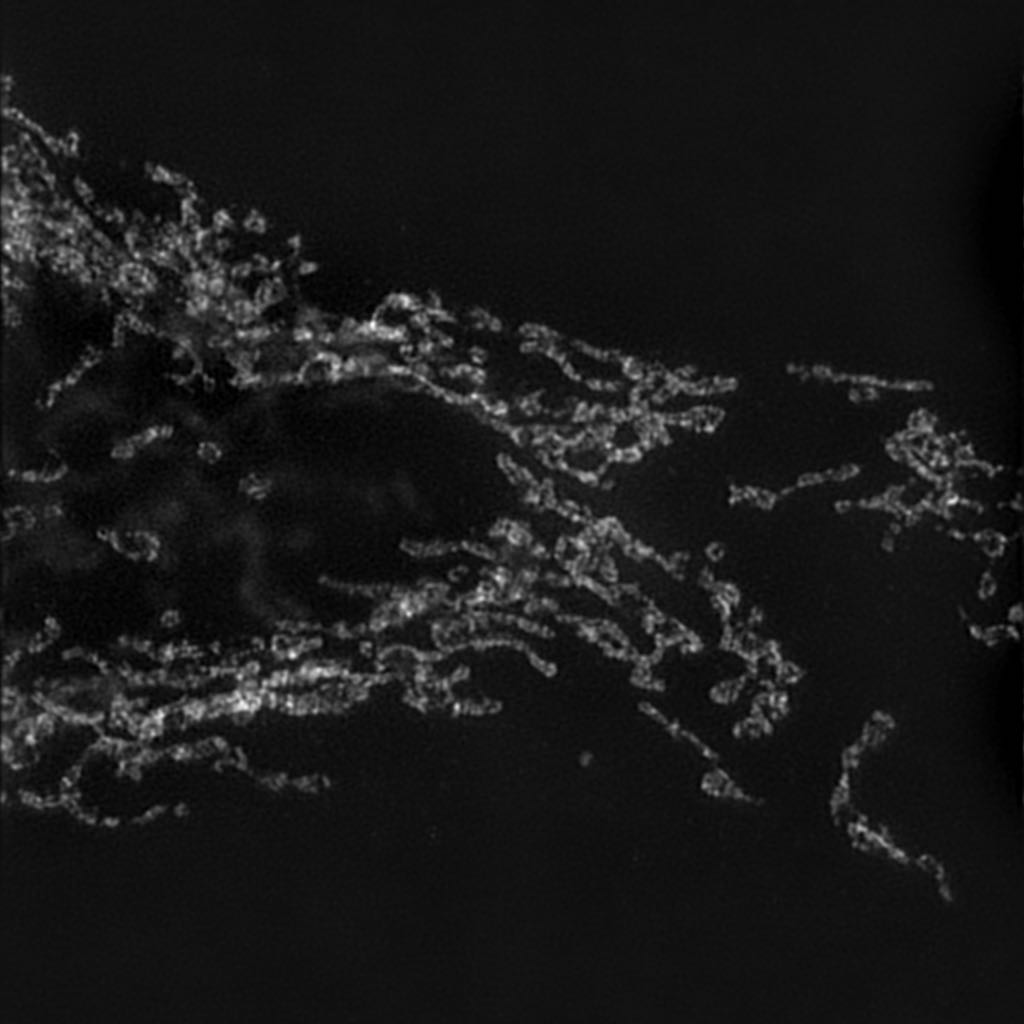}

High resolution (target)
\end{minipage}
\caption{Image pair from W2S \cite{W2S}. The low resolution image has half the size of the high resolution one. For better visualisation, it is upsampled by a factor 2 using bicubic interpolation}
\end{figure}

One recent approach to improve the performance of denoising and super-resolution algorithms, though addressed individually, is Stochastic Frequency Masking (SFM)~\cite{SFM}. It essentially consists of an image manipulation technique using masking in the frequency domain that can be applied during the training of any learning method with little additional cost. This masking alters the underlying training of the restoration networks, in accordance with a conditional learning framework presented by the authors. SFM empirically improves the performance of the tested state-of-the-art algorithms, individually on denoising and on super-resolution tasks. In this project, we address the denoising and super-resolution problems jointly, on the W2S dataset. SFM could be adapted for the joint scenario, but this is outside the scope of this project.

For the model and training pipeline, we choose the framework developed by Wang \etal in ``Enhanced Super-Resolution Generative Adversarial Networks'' (ESRGAN) \cite{ESRGAN}, which was declared during the PIRM2018 challenge on super-resolution to have the best perceptual quality among super resolution algorithms, both by numerical metrics and human assessment \cite{PIRM-2018}. The use of GANs also remains widespread in recent super-resolution challenges~\cite{AIM-2019}.

The project consists in implementing ESRGAN in Python, using the TensorFlow library \cite{tensorflow} on the W2S\cite{W2S} dataset. We experiment different losses, and output our best method as a pretrained model that can be used for inference and will be added to the DeepImageJ \cite{DeepImageJ} plugin for ImageJ \cite{ImageJ}.

\section{Implementation}
This section describes the implementation of ESRGAN \cite{ESRGAN}, as well as the design choices we make for this project.

\subsection{Data pre-processing}
The first step of the data preparation is to load the dataset and break down the images into patches. In this project, we choose to use patches of 64 $\times$ 64 pixels for low resolution images and of 128 $\times$ 128 pixels in high resolution. This allows to do smaller but quicker steps during the learning, compared to processing entire images. The patches are built with 50\% overlap. We hence extract $15 \times 15=225$ patches per image. These patches can either be created once, saved on the hard drive, and then loaded at the start of each training, or can be built directly at the start of a training. After testing both approaches, we decide to keep the online method as it is not slower than loading the extracted patches and does not require extra disk storage space.

The dataset is then shuffled. While being a simple procedure, a challenge arises, as the low resolution (LR) and high resolution (HR) patch pairs have to stay coherent. This is handled by considering the pair as a tuple and by applying the same operations on both sides of the pair at every step of the processing pipeline. When shuffling, the tuples are shuffled instead of single patches.

We also augment the dataset to make it vary between each epoch. Every patch pair has 50\% chance of being transformed and if so, is randomly flipped horizontally, vertically and rotated by 90\si{\degree} with 50\% chance for every transformation. That means that every patch is either unchanged or randomly transformed into one of eight configurations. This is particularly important to avoid overfitting.

Finally, the patches are regrouped in batches and pre-fetched to the computation device(s).

\subsection{Generator} \label{generator}
\begin{figure}[h]
\includegraphics[width=\linewidth]{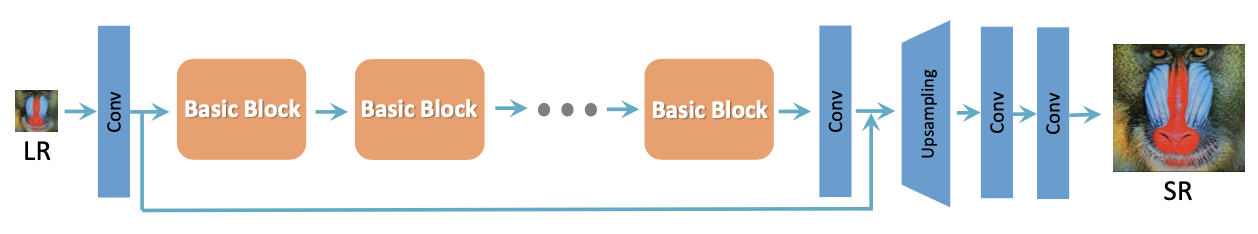}
\caption{Architecture of the residual network. Figure taken from ESRGAN
\cite{ESRGAN}.}\label{fig:RRDBNet}
\end{figure}

The generator is the network responsible for the processing and enhancement of the images. It is fed with low resolution images and outputs a denoised image of twice the resolution. It also serves as our inference model. Our network has the same architecture as ESRGAN \cite{ESRGAN} (see Figure \ref{fig:RRDBNet}, that is based on the SSRNet residual network introduced by Ledig \etal in ``SRGAN'' \cite{SRGAN}.

\begin{figure}[h]
\includegraphics[width=\linewidth]{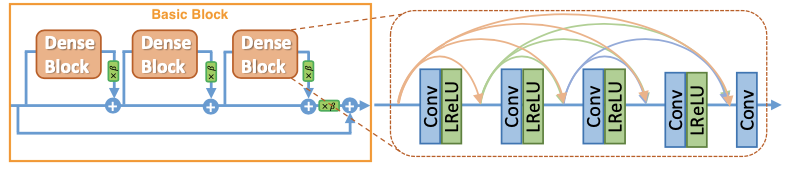}
\caption{Architecture of one Basic Block (RRDB) of the network. Figure modified from ESRGAN \cite{ESRGAN}.}\label{fig:RRDB}
\end{figure}

The idea is to perform most of the computation in the LR space, preceding an up-sampling layer and two final convolution layers in the HR domain. The computation trunk is a sequence of simpler elements, called Residual-in-Residual Dense Blocks (RRDBs) that all have shortcut and residual connections (see Figure \ref{fig:RRDB}).

We implement this model using TensorFlow with the model subclassing API, that allows custom layers and activation. We define a model class for both the Basic Blocks and Dense Blocks and then are able to consider them as simple layers at higher abstraction levels. The convolution kernels are initialized using a scaled down version of the MSRA initializer by He \etal \cite{MSRA}. It was shown in ESRGAN \cite{ESRGAN} that such scaled down parameter can increase generator stability during training, especially for big networks, with which the use of a bad initialization can lead to inconsistent image brightness.

\subsection{Discriminator} \label{Discriminator}

The discriminator is the adversarial counterpart of the generator. Its only purpose is to learn to distinguish a real image from a generated one. In our case, to be able to identify an image as being from the HR dataset or a super-resolution image outputted by the model. The discriminator is trained side by side with the generator, which will, as it becomes better at enhancing images, try to trick the discriminator, which will also improve its predictions by learning from its mistakes and correct guesses.

As was shown by Mahendran and Vedaldi \cite{DeepImageRepresentations}, layers of image classification networks each contain a representation of the images, that becomes more abstract as we progress through the network. That means that  such a network (for example VGG \cite{VGG} networks) retains information on the content and representation of the image. It thus seems plausible that a similar architecture can be used to effectively identify an input image as real or fake, through a simple classification task.

We use the same network as the discriminator developed by Ledig \etal in ``SRGAN'' \cite{SRGAN}, that follows the architecture of the VGG \cite{VGG} network, by using strided convolutions to extract features from the image, and using two dense layers and a sigmoid activation to output a probability for the image of being real or fake. 

\begin{figure}[h]
\includegraphics[width=\linewidth]{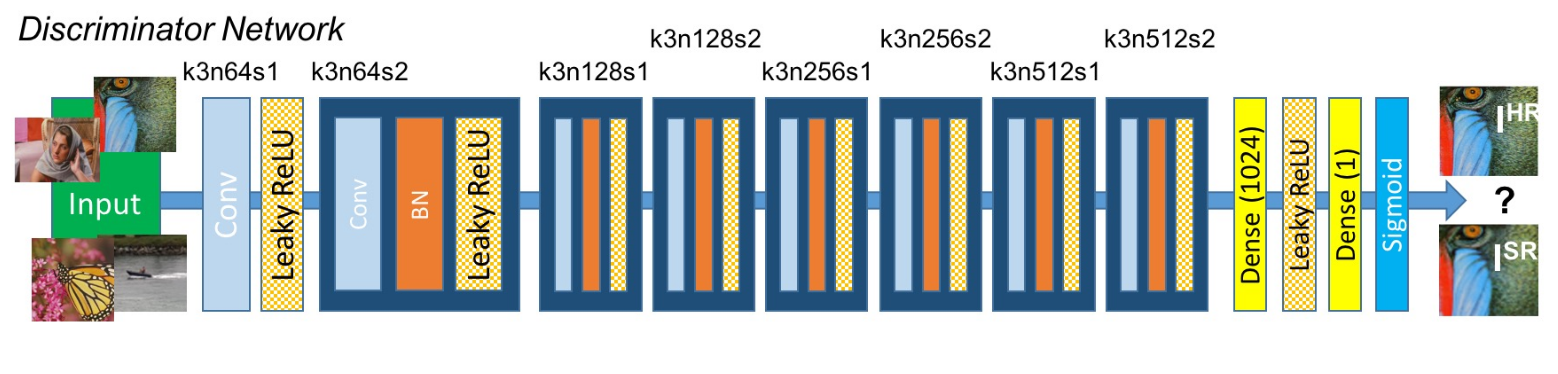}
\caption{Architecture of discriminator network in SRGAN. Figure from SRGAN \cite{SRGAN}.} \label{fig:Disc}
\end{figure}

As in ESRGAN \cite{ESRGAN}, we transform this network into a Relativistic Discriminator \cite{RaD}, by considering the output before the sigmoid activation and computing the ``probability that a real image $x_r$ is relatively more realistic than a fake one $x_f$'' \cite{ESRGAN} by comparing the activation of real high-resolution patches to the mean activation of the fake generated patches of the batch, and the activation of fake patches to the mean activation of real patches. This procedure will be detailed in section \ref{loss function}.

The implementation of the discriminator network in TensorFlow is straightforward and
follows the model description from Figure \ref{fig:Disc}, without the final sigmoid activation. The network is implemented using the model subclassing API, as the generator.

\subsection{Loss functions} \label{loss function}

As in ESRGAN \cite{ESRGAN}, the generator loss is an aggregation of three components. There is first a pixel-wise loss component that ensures that the image does not drift away from the target, at the pixel level. Second is a perceptual loss component, that brings the image closer to the target at the perceptual (feature) level and third is the adversarial component that comes from the discriminator, that grades the realism of the output image.

The discriminator loss is only composed of an adversarial part and will be covered in section \ref{adv}.

\subsubsection{Pixel-wise loss} \label{pixel-wise loss}

The pixel wise loss aims at making each pixel of the output image match as close as possible the pixels of the target image. It is most of the time implemented as a a $\ell$1 or squared $\ell$2 norm of the subtraction of image X and Y of size M$\times$N, as
\begin{gather*}
\ell1(X,Y) = \frac{1}{M \times N} \times \sum_{i=1}^{M} \sum_{j=1}^{N} \abs{x_{ij} - y_{ij}} \\
\ell2(X,Y) = \frac{1}{M \times N} \times \sum_{i=1}^{M} \sum_{j=1}^{N} \abs{x_{ij} - y_{ij}}^2
\end{gather*}

This method aims at optimizing the Peak Signal-to-Noise Ratio (PSNR) of the approximation, directly when using $\ell$2 loss, as the PSNR is computed with the mean squared error, or indirectly using the $\ell$1 loss, as it also minimizes the pixel difference.

Using only a pixel-wise loss for optimization already provides some degree of denoising. However, as shown by Ledig \etal \cite{SRGAN}, a high PSNR value (low pixel-wise loss) does not necessarily mean images of better visual quality, as this loss does not capture differences in textures, structures or levels of detail. This is the reason we use the perceptual loss to capture such features.

\subsubsection{Perceptual loss}

The use of perceptual loss aims at reproducing the way humans recognize images, by looking at its content, structure, shapes and colors instead of pixel-by-pixel comparison. As was stated in section \ref{Discriminator}, an image classification network as VGG \cite{VGG} does carry such features when processing an image. Johnson \etal \cite{percept} introduced the use of the output of an intermediate layer of a VGG network trained on image classification. This output carries high-level features. We can hence try to compute a perceptual similarity (or difference) of our generated image and the groundtruth image by feeding them both to the network and extracting their representation at an intermediate layer. Johnson \etal defined the perceptual loss as the pixel-wise loss on the feature map of the images. 

As in ESRGAN \cite{ESRGAN}, we use a VGG-19 network that is pre-trained on the ImageNet \cite{ImageNet} dataset, that consists of five convolutional blocks. As taking the output at the very end of the network would yield too abstract features and would not retain the details of the image, as extracting the features from the very first layers would not be very different from our previous pixel-wise loss. This is why we choose to extract the loss from a layer at the middle of the network, after the last convolutional layer of the third block. As proposed in ESRGAN \cite{ESRGAN}, the output is extracted before the activation layer, as it carries more information and better prevents brightness alteration.

The model is implemented in the code by importing the pre-trained VGG network from the TensorFlow's Keras applications module, and is adapted to have the input shape of our patch size and truncated at the end of the third convolutional block.

\subsection{Adversarial loss} \label{adv}

The adversarial part allows to go further than perceptual loss by adding some degree of adaptability given that the discriminator also learns from its inputs and predictions, that the image classifier used for perceptual loss has frozen weights. The adversarial loss has the role of rewarding the generator for images that tricked the discriminator, which is penalized, and vice versa. 

The relativistic adversarial loss we implement follows the one of ESRGAN. For each batch of patches, we first compute the discriminator output (feature map) C(x) of each image x and the mean feature map of each class (generated/HR) in the batch, $\mathbb{E}[x_{f}]$ and $\mathbb{E}[x_{r}]$. The Relativistic average Discriminator RaD \cite{RaD} for a generated/HR pair $x_{f}, x_{r}$ , denoted $D_{Ra}$ (for notation consistency with ESRGAN), is defined as:
\begin{gather*}
D_{Ra}(x_{r}, x_{f}) = \sigma(C(x_{r}) - \mathbb{E}[x_{f}]) \\
D_{Ra}(x_{f}, x_{r}) = \sigma(C(x_{f}) - \mathbb{E}[x_{r}])
\end{gather*}
where $\sigma$ is the sigmoid function.

The generator loss $L_{G}^{Ra}$ and discriminator loss $L_{D}^{Ra}$ are then defined as:

\begin{align*}
L_{G}^{Ra} = &-\mathbb{E}_{x_{r}}[\log(1-D_{Ra}(x_{r}, x_{f}))]\\
&-\mathbb{E}_{x_{f}}[\log(D_{Ra}(x_{f}, x_{r}))] \\
\\
L_{D}^{Ra} = &-\mathbb{E}_{x_{r}}[\log(D_{Ra}(x_{r}, x_{f}))]\\ &-\mathbb{E}_{x_{f}}[\log(1-D_{Ra}(x_{f}, x_{r}))]
\end{align*}

In TensorFlow, using this exact formula yields numerically unstable results. The workaround used in the code consists in expressing each side of both losses as a one-sided binary cross-entropy loss, which means both the sigmoid and log function stability is handled by the TensorFlow function.

\subsection{Optimizers}

The optimizer for the generator and discriminator is an Adam optimizer \cite{Adam}, for which we use the AMSGrad variant proposed by Reddi \etal \cite{AMSGrad}, which fixes the convergence issues of the original Adam optimizer and can improve performance. As in ESRGAN, the learning rate decays over time. We choose to divide the learning rate by a factor of 2 between each epoch to help avoiding overfit.

The training is performed in two phases. The first consists in training the generator on the images using only pixel-wise loss while the second includes pixel-wise loss, perceptual loss and adversarial loss using a discriminator that is trained at the same time. This first training phase produces an approximation that serves as a warm-up for the generator optimization, which leads it in the correct direction. In addition, the discriminator starts its training with well approximated images, which allows it to directly focus on observing the features instead of having to classify trivially wrong images, as would be the output of an untrained generator.

\section{Results}

In this section, we present our final results, as well as the different steps leading to this solution and other tested experiments. Note that none of the presented images were seen by the network during training. We compute the PSNR and structural similarity index (SSIM) of the generated images compared to the ground-truth for a numerical performance, but the quality of the results are best visually assessed  given that, as stated in section \ref{pixel-wise loss}, pixel-wise scoring metrics disagree with perceptual quality of the reconstruction, which is our aim in this project.

\subsection{Setup}

All experiments were realized with a generator with 23 RRDB blocks (see Figure \ref{fig:RRDB}, with a residual factor of 0.2 and activation layers with a leak factor of 0.2, weights initialized with MSRA\cite{MSRA} scaled down by a factor 10. The networks are optimized using Adam \cite{Adam} with parameters $\beta_{1} = 0.9$ and $\beta_{2} = 0.99$. The learning rate is initialized to $1 \times 10^{-4}$ and divided by two after each epoch for 50 epochs (for both phases of training). The experiments were run on two NVIDIA Titan X GPUs.

\begin{figure}[H]
\centering

\begin{subfigure}{0.24\textwidth}
\caption{Bicubic \\ (20.35 / 0.41)}
\includegraphics[width=\linewidth]{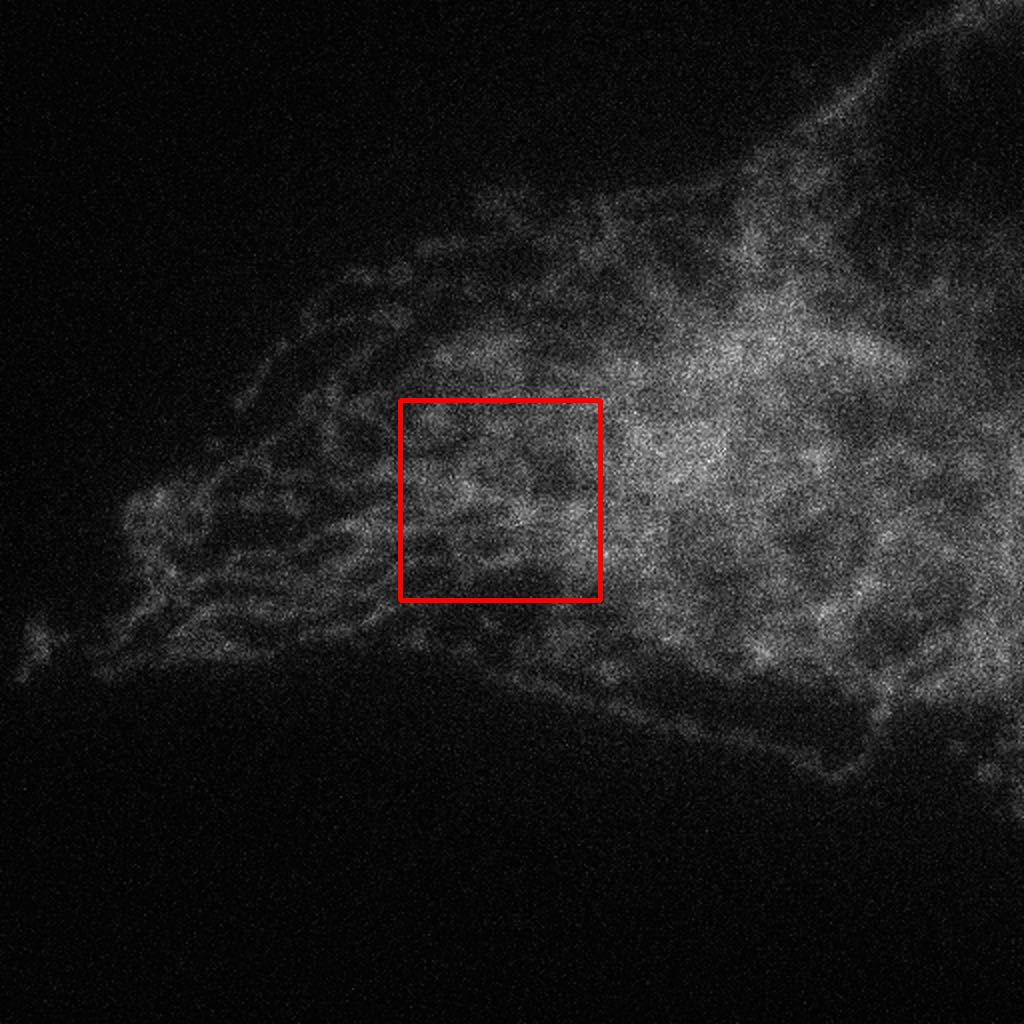}
\end{subfigure}\hspace*{\fill}
\begin{subfigure}{0.24\textwidth}
\caption{Perceptual loss \\ (24.64 / 0.71)}
\includegraphics[width=\linewidth]{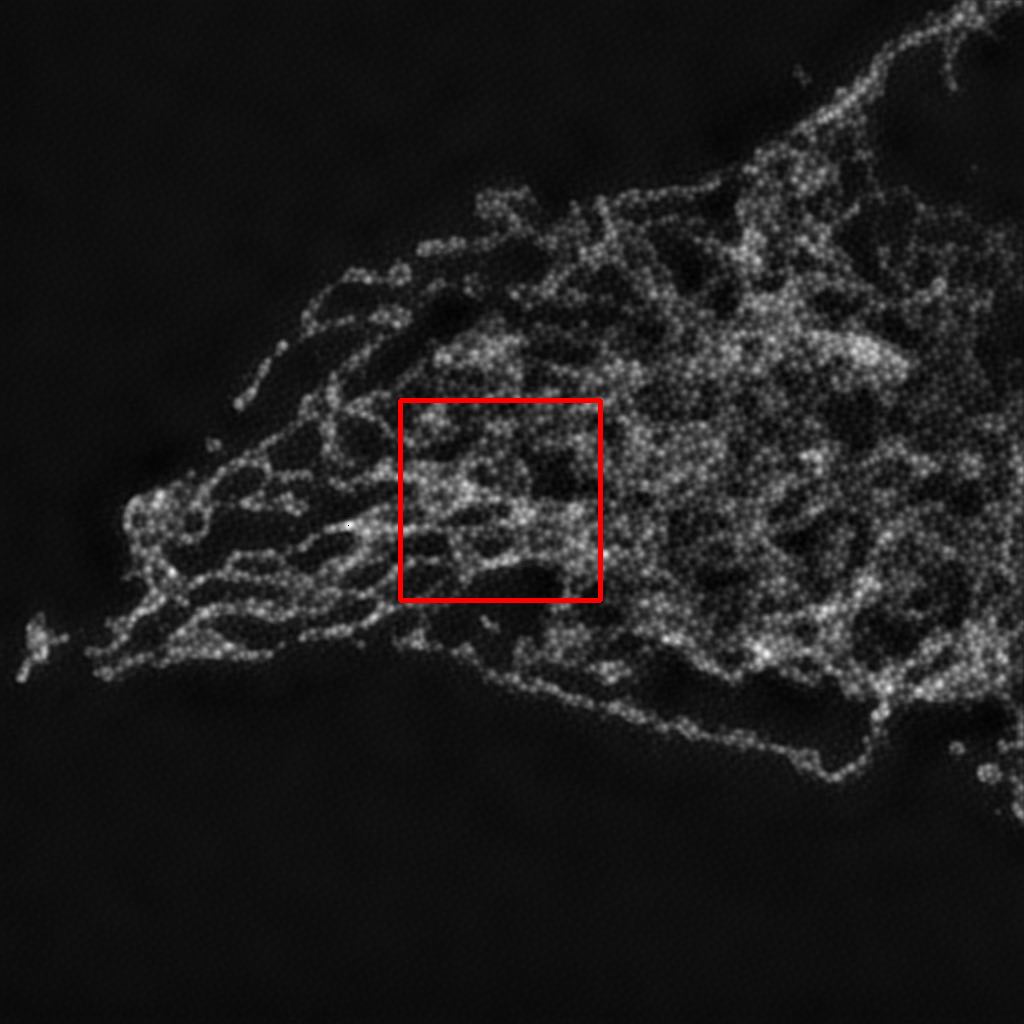}
\end{subfigure}\hspace*{\fill}
\begin{subfigure}{0.24\textwidth}
\caption{Texture loss \\ (24.78 / 0.77)}
\includegraphics[width=\linewidth]{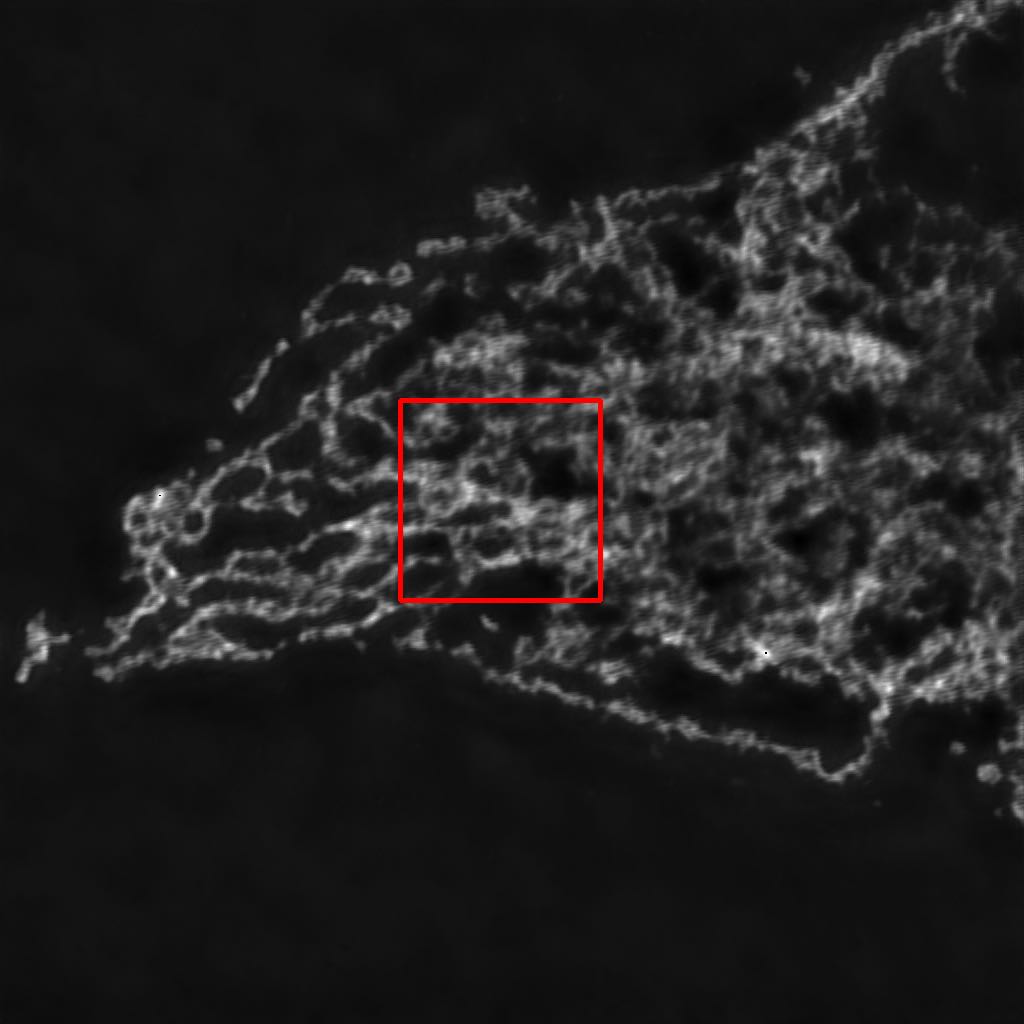}
\end{subfigure}\hspace*{\fill}
\begin{subfigure}{0.24\textwidth}
\caption{SIM \\ \phantom{a}}
\includegraphics[width=\linewidth]{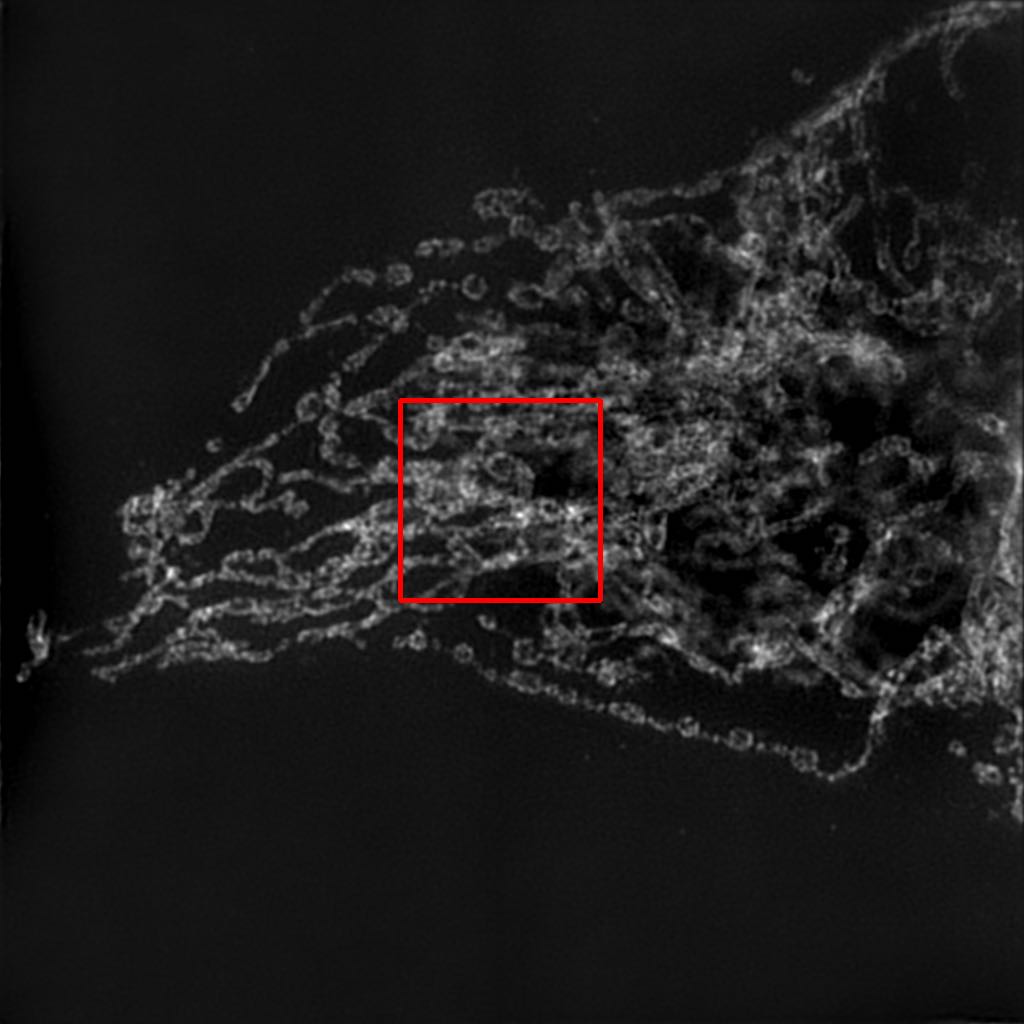}
\end{subfigure}

\medskip

\begin{subfigure}{0.24\textwidth}
\includegraphics[width=\linewidth]{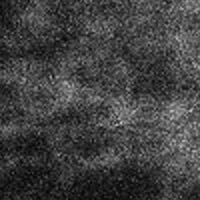}
\end{subfigure}\hspace*{\fill}
\begin{subfigure}{0.24\textwidth}
\includegraphics[width=\linewidth]{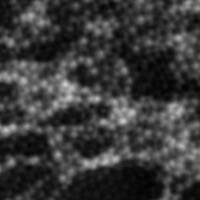}
\end{subfigure}\hspace*{\fill}
\begin{subfigure}{0.24\textwidth}
\includegraphics[width=\linewidth]{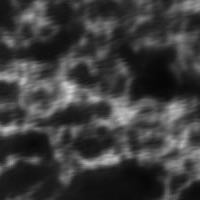}
\end{subfigure}\hspace*{\fill}
\begin{subfigure}{0.24\textwidth}
\includegraphics[width=\linewidth]{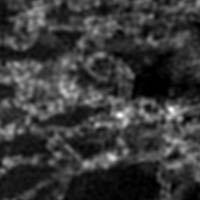}
\end{subfigure}

\caption{Comparison of the use of perceptual and texture loss. (a) Low resolution widefield image upscaled using bicubic interpolation. (b) Output of our model after training with pixel-wise and perceptual loss. (c) Output of our model after training with pixel-wise and texture loss. (d) Corresponding image obtained using a SIM process. PSNR and SSIM values are shown in parenthesis.} \label{fig:gram}
\end{figure}

\subsection{Experiments}

\subsubsection{Experiment on different loss}

During the project, we experimented the use of texture loss, introduced by Gatys \etal \cite{texture}, and used in EnhanceNet \cite{enhancenet}, that computes the feature correlation between two images, using the gram matrix of the feature maps at the extracted layer and then applies pixel-wise loss on those matrices. This can be seen as a second order perceptual loss. To test this method, we compare it to the perceptual loss using pixel-wise loss and perceptual/texture loss respectively.

We can observe in Figure \ref{fig:gram} that the image obtained with texture loss (panel c) is very blurry and contains some artifacts in the background. These poor results can be explained by the fact that, as explained in Enhancenet \cite{enhancenet}, this technique works best if the target texture is provided at test time. This approach is more suitable for applications as style transfer \cite{style}, where the texture target is known.

\subsubsection{Experiment on L2 loss}

We also experimented the use of $\ell$2 loss for the pixel and perceptual loss. 

\begin{figure}[H]
\centering

\begin{subfigure}{0.24\textwidth}
\caption{Bicubic \\ (20.35 / 0.41)}
\includegraphics[width=\linewidth]{images/res/lr.jpg}
\end{subfigure}\hspace*{\fill}
\begin{subfigure}{0.24\textwidth}
\caption{$\ell$1 pixel-wise loss \\ (25.44 / 0.77)}
\includegraphics[width=\linewidth]{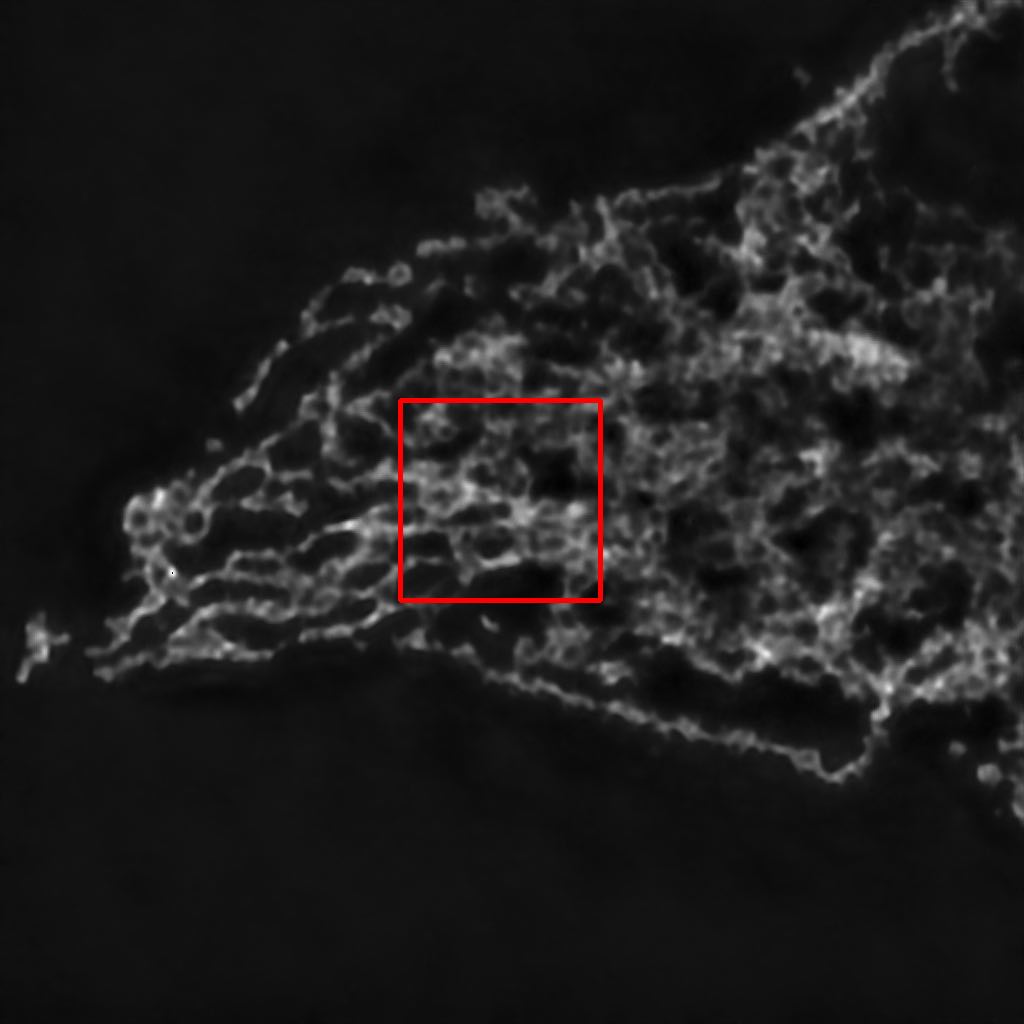}
\end{subfigure}\hspace*{\fill}
\begin{subfigure}{0.24\textwidth}
\caption{$\ell$2 pixel-wise loss \\ (24.23 / 0.76)}
\includegraphics[width=\linewidth]{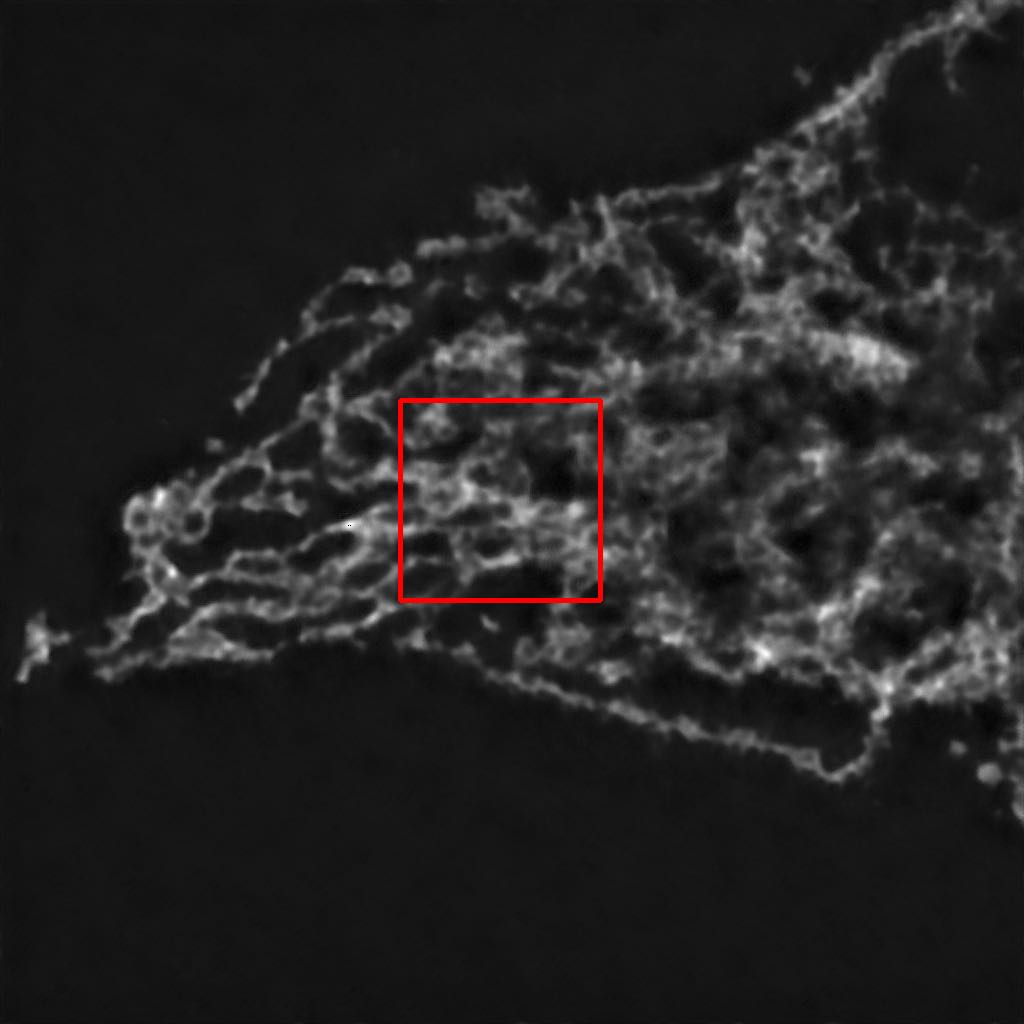}
\end{subfigure}\hspace*{\fill}
\begin{subfigure}{0.24\textwidth}
\caption{SIM \\ \phantom{a}}
\includegraphics[width=\linewidth]{images/res/hr.jpg}
\end{subfigure}

\medskip

\begin{subfigure}{0.24\textwidth}
\includegraphics[width=\linewidth]{images/res/lr_patch.jpg}
\end{subfigure}\hspace*{\fill}
\begin{subfigure}{0.24\textwidth}
\includegraphics[width=\linewidth]{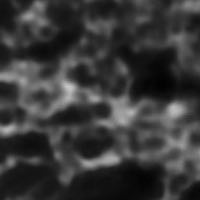}
\end{subfigure}\hspace*{\fill}
\begin{subfigure}{0.24\textwidth}
\includegraphics[width=\linewidth]{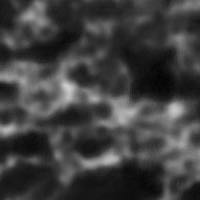}
\end{subfigure}\hspace*{\fill}
\begin{subfigure}{0.24\textwidth}
\includegraphics[width=\linewidth]{images/res/hr_patch.jpg}
\end{subfigure}

\caption{Comparison of the use of $\ell$1 and $\ell$2 loss as our pixel-wise loss for pre-training. (a) Low resolution widefield image upscaled using bicubic interpolation, (b) Output of our model after the first training phase using pixel-wise $\ell$1 loss only. (c) Output of our model after the first training phase using pixel-wise $\ell$2 loss only. (d) Corresponding image obtained using a SIM process. PSNR and SSIM values are shown in parenthesis.} \label{fig:l1l2_pix}
\end{figure}

\begin{figure}[H]
\centering

\begin{subfigure}{0.24\textwidth}
\caption{Bicubic \\ (20.35 / 0.41)}
\includegraphics[width=\linewidth]{images/res/lr.jpg}
\end{subfigure}\hspace*{\fill}
\begin{subfigure}{0.24\textwidth}
\caption{$\ell$1 GAN loss \\ (24.96 / 0.74)}
\includegraphics[width=\linewidth]{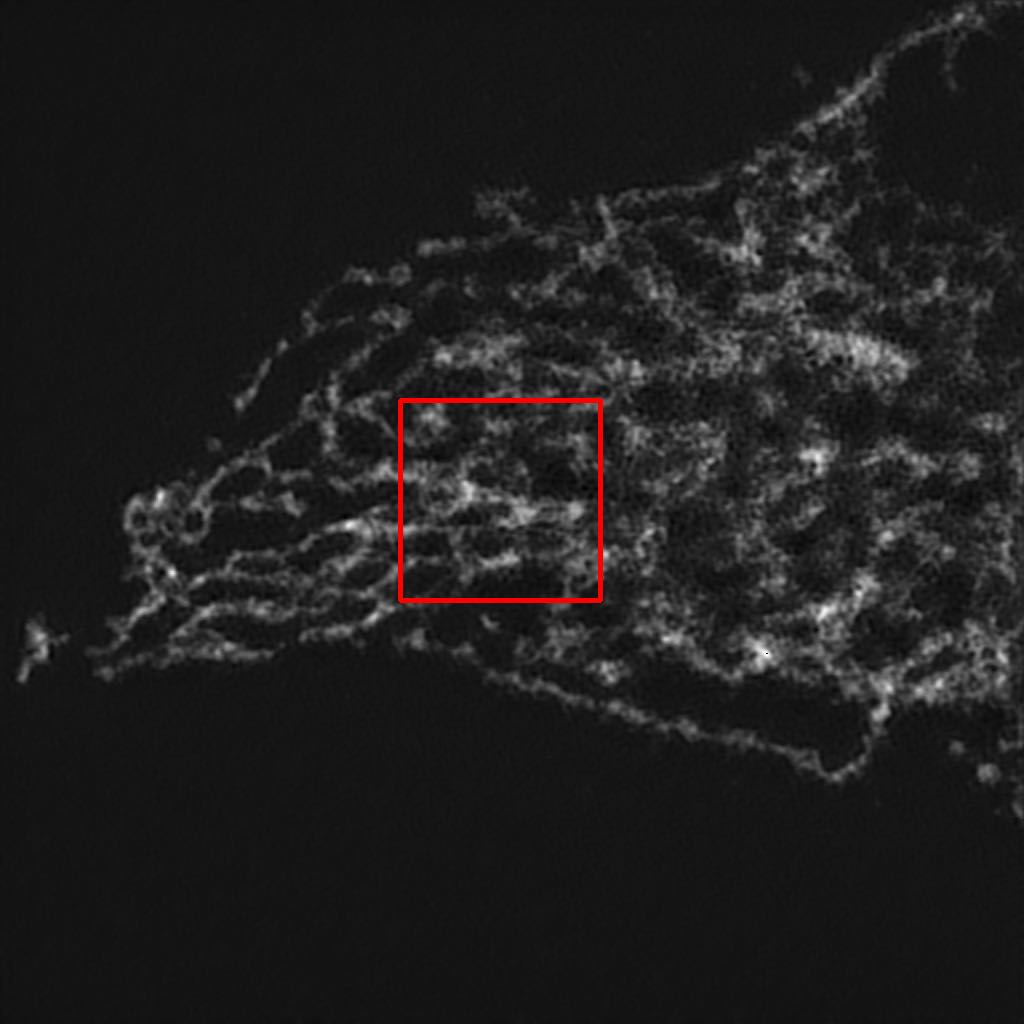}
\end{subfigure}\hspace*{\fill}
\begin{subfigure}{0.24\textwidth}
\caption{$\ell$2 GAN loss \\ (16.60 / 0.48)}
\includegraphics[width=\linewidth]{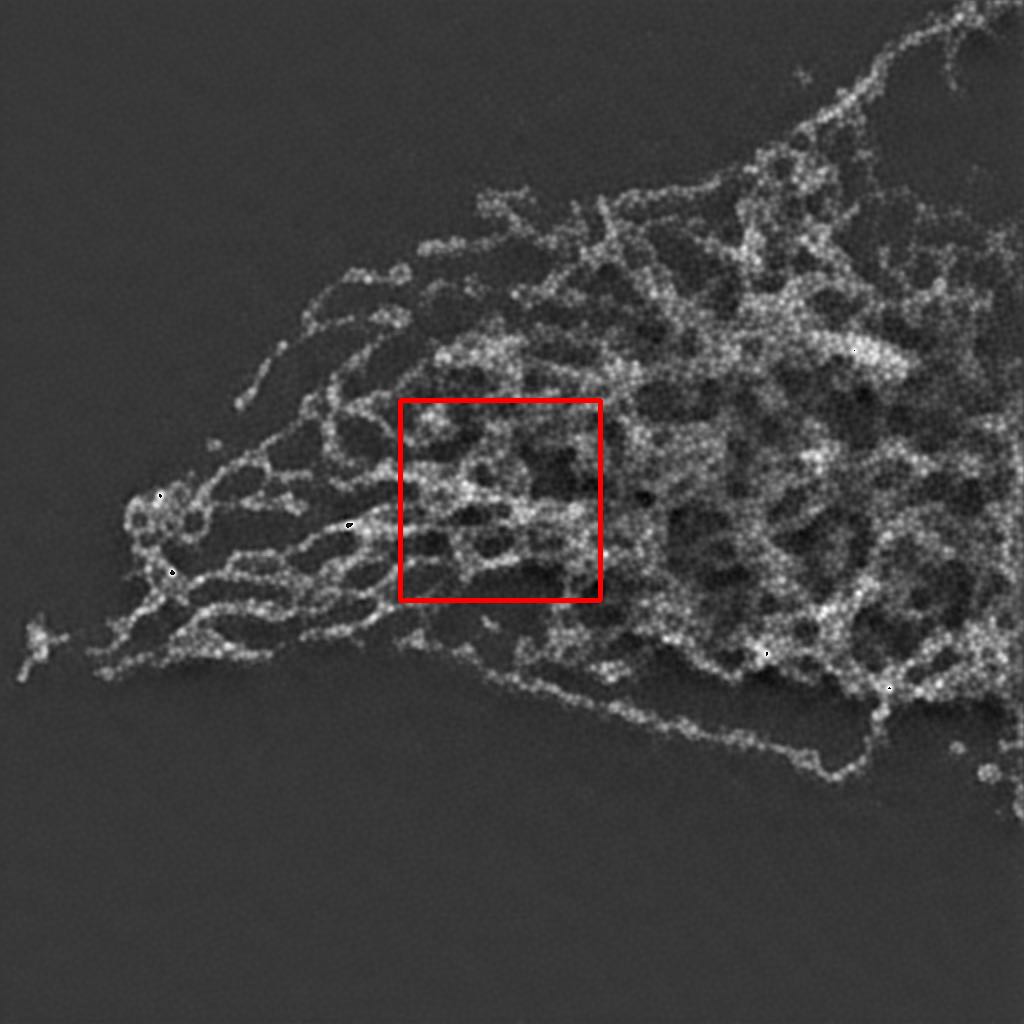}
\end{subfigure}\hspace*{\fill}
\begin{subfigure}{0.24\textwidth}
\caption{SIM \\ \phantom{a}}
\includegraphics[width=\linewidth]{images/res/hr.jpg}
\end{subfigure}

\medskip

\begin{subfigure}{0.24\textwidth}
\includegraphics[width=\linewidth]{images/res/lr_patch.jpg}
\end{subfigure}\hspace*{\fill}
\begin{subfigure}{0.24\textwidth}
\includegraphics[width=\linewidth]{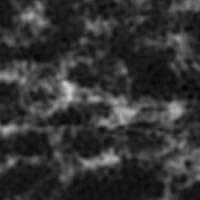}
\end{subfigure}\hspace*{\fill}
\begin{subfigure}{0.24\textwidth}
\includegraphics[width=\linewidth]{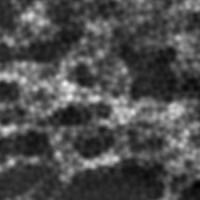}
\end{subfigure}\hspace*{\fill}
\begin{subfigure}{0.24\textwidth}
\includegraphics[width=\linewidth]{images/res/hr_patch.jpg}
\end{subfigure}

\caption{Comparison of the use of $\ell$1 and $\ell$2 loss for both pixel-wise and perceptual loss for the full training. (a) Low resolution widefield image upscaled using bicubic interpolation, (b) Output of our model after the full training phase using $\ell$1 loss for pixel-wise and perceptual loss. (c) Output of our model after the full training phase using $\ell$2 loss for pixel-wise and perceptual loss. (d) Corresponding image obtained using a SIM process. PSNR and SSIM values are shown in parenthesis.} \label{fig:l1l2_gan}
\end{figure}

It can be observed in Figure \ref{fig:l1l2_pix} that for the pixel-wise loss, the use of the $\ell$2 loss (panel c) results in blurrier images. In Figure \ref{fig:l1l2_gan}, the use of the $\ell$2 loss (panel c) is unstable and leads to an increase in the image brightness. On the image of Figure \ref{fig:l1l2_pix}, the mean pixel value using the $\ell$2 loss is twice the value of the other. This can relate to brightness instability observed with a too large initialisation, as explained in section \ref{generator}. We hence choose the $\ell$1 criterion for the aggregation of our losses.

\subsection{Final results} \label{final_res}

Our final results were obtained by using an $\ell$1 loss for both pixel-wise loss and perceptual loss, with weights of $1\times10^{-2}$ for pixel-wise loss during pre-training and training, and for the perceptual loss during main training. As the value of the adversarial loss is low, we use a weight of 1.0 for both the optimizer and the discriminator.

\begin{figure}[H]
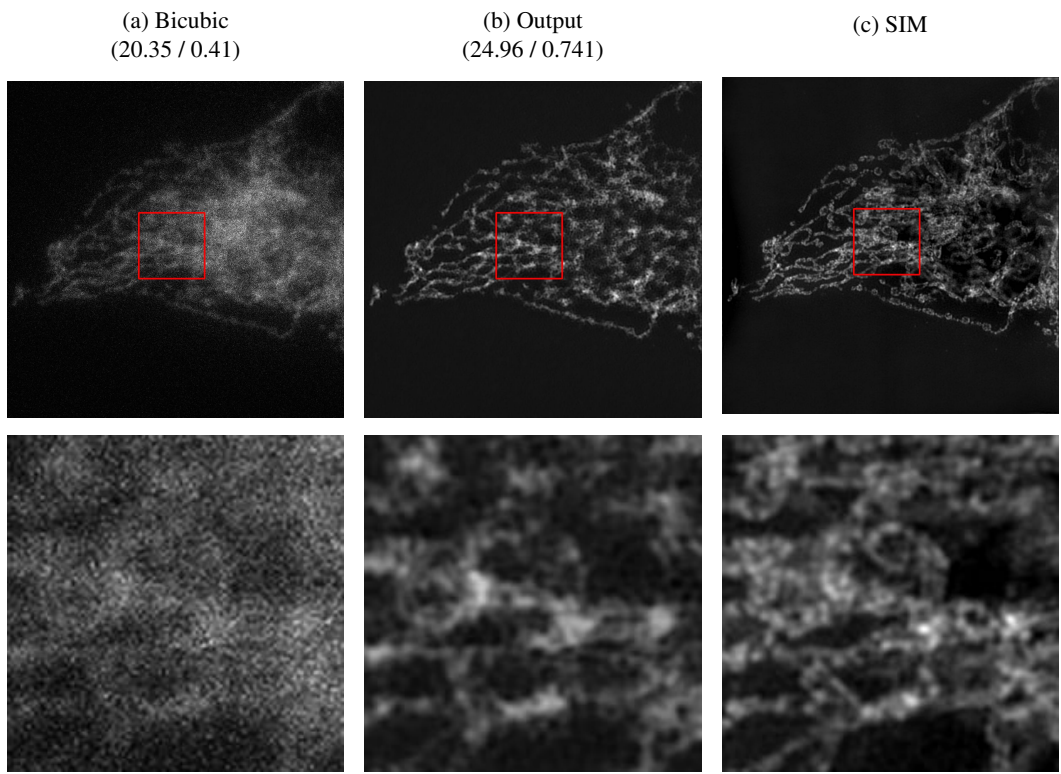

\centering
\begin{subfigure}{0.32\textwidth}
\caption{Bicubic \\ (20.35 / 0.41)}
\includegraphics[width=\linewidth]{images/res/lr.jpg}
\end{subfigure}\hspace*{\fill}
\begin{subfigure}{0.32\textwidth}
\caption{Output \\ (24.96 / 0.741)}
\includegraphics[width=\linewidth]{images/res/gan.jpg}
\end{subfigure}\hspace*{\fill}
\begin{subfigure}{0.32\textwidth}
\caption{SIM \\ \phantom{a} }
\includegraphics[width=\linewidth]{images/res/hr.jpg}
\end{subfigure}

\medskip

\begin{subfigure}{0.32\textwidth}
\includegraphics[width=\linewidth]{images/res/lr_patch.jpg}
\end{subfigure}\hspace*{\fill}
\begin{subfigure}{0.32\textwidth}
\includegraphics[width=\linewidth]{images/res/gan_patch.jpg}
\end{subfigure}\hspace*{\fill}
\begin{subfigure}{0.32\textwidth}
\includegraphics[width=\linewidth]{images/res/hr_patch.jpg}
\end{subfigure}

\caption{Results of our algorithm on an unseen image in the W2S dataset \cite{W2S}. (a) Low resolution widefield image upscaled using bicubic interpolation. (b) Output of our model in our final solution. (c) Corresponding image obtained using a SIM process. PSNR and SSIM values are, respectively, shown in parenthesis. Images on the second row are zoomed up regions highlighted by red boxes on the full images.} \label{fig:final_res}
\end{figure}

It can be observed in Figure \ref{fig:final_res} that, even though not perfect, the denoising is very effective and that the textures and details of the image are well approximated when compared to the ground truth SIM image.

For completeness, we also show the intermediate outputs of the algorithm, using the same parameters as in section \ref{final_res}. The first of the generated images (panel b) is the output of the network after the pre-training phase using pixel-wise loss only. The second (panel c) is after a a full training phase using pixel-wise and perceptual loss (no adversarial part) and the last one (panel d) is with the full loss. The results are shown for the same cell sample on three different channels.

\begin{figure}[H]
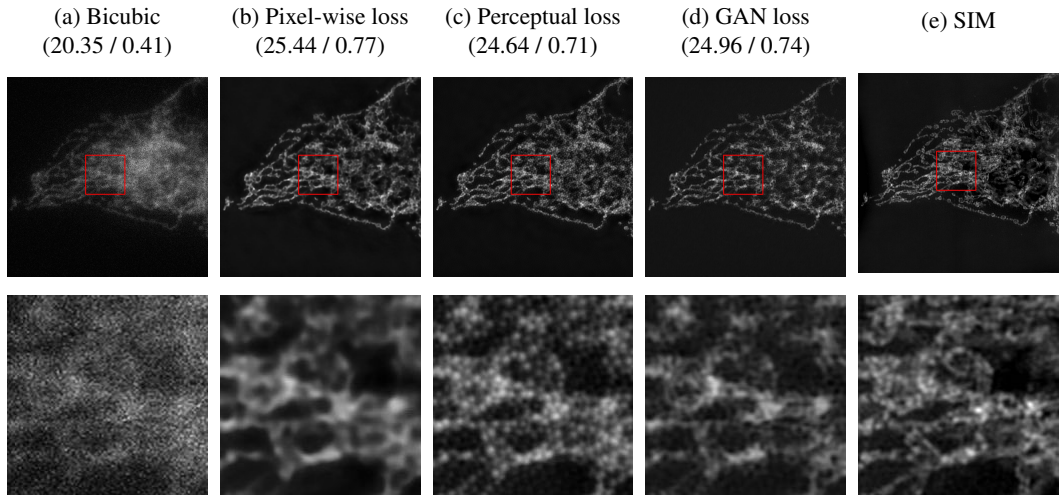

\centering

\begin{subfigure}{0.19\textwidth}
\caption{Bicubic \\ (20.35 / 0.41)}
\includegraphics[width=\linewidth]{images/res/lr.jpg}
\end{subfigure}\hspace*{\fill}
\begin{subfigure}{0.19\textwidth}
\caption{Pixel-wise loss \\ (25.44 / 0.77)}
\includegraphics[width=\linewidth]{images/res/pix.jpg}
\end{subfigure}\hspace*{\fill}
\begin{subfigure}{0.19\textwidth}
\caption{Perceptual loss \\ (24.64 / 0.71)}
\includegraphics[width=\linewidth]{images/res/feat.jpg}
\end{subfigure}\hspace*{\fill}
\begin{subfigure}{0.19\textwidth}
\caption{GAN loss \\ (24.96 / 0.74)}
\includegraphics[width=\linewidth]{images/res/gan.jpg}
\end{subfigure}\hspace*{\fill}
\begin{subfigure}{0.19\textwidth}
\caption{SIM \\ \phantom{a}}
\includegraphics[width=\linewidth]{images/res/hr.jpg}
\end{subfigure}

\medskip

\begin{subfigure}{0.19\textwidth}
\includegraphics[width=\linewidth]{images/res/lr_patch.jpg}
\end{subfigure}\hspace*{\fill}
\begin{subfigure}{0.19\textwidth}
\includegraphics[width=\linewidth]{images/res/pix_patch.jpg}
\end{subfigure}\hspace*{\fill}
\begin{subfigure}{0.19\textwidth}
\includegraphics[width=\linewidth]{images/res/feat_patch.jpg}
\end{subfigure}\hspace*{\fill}
\begin{subfigure}{0.19\textwidth}
\includegraphics[width=\linewidth]{images/res/gan_patch.jpg}
\end{subfigure}\hspace*{\fill}
\begin{subfigure}{0.19\textwidth}
\includegraphics[width=\linewidth]{images/res/hr_patch.jpg}
\end{subfigure}

\caption{Results of our algorithm on an unseen image in the W2S dataset \cite{W2S}, first channel. (a) Low resolution widefield image upscaled using bicubic interpolation. (b) Output of our model after the first training phase using pixel-wise loss only. (c) Output of our model after the full training using pixel-wise and perceptual loss.(d) Output of our model in our final solution using full loss. (e) Corresponding image obtained using a SIM process. PSNR and SSIM values are shown in parenthesis.} \label{fig:inter_c1}
\end{figure}

\begin{figure}[H]
\centering

\begin{subfigure}{0.19\textwidth}
\caption{Bicubic \\ (16.55 / 0.27)}
\includegraphics[width=\linewidth]{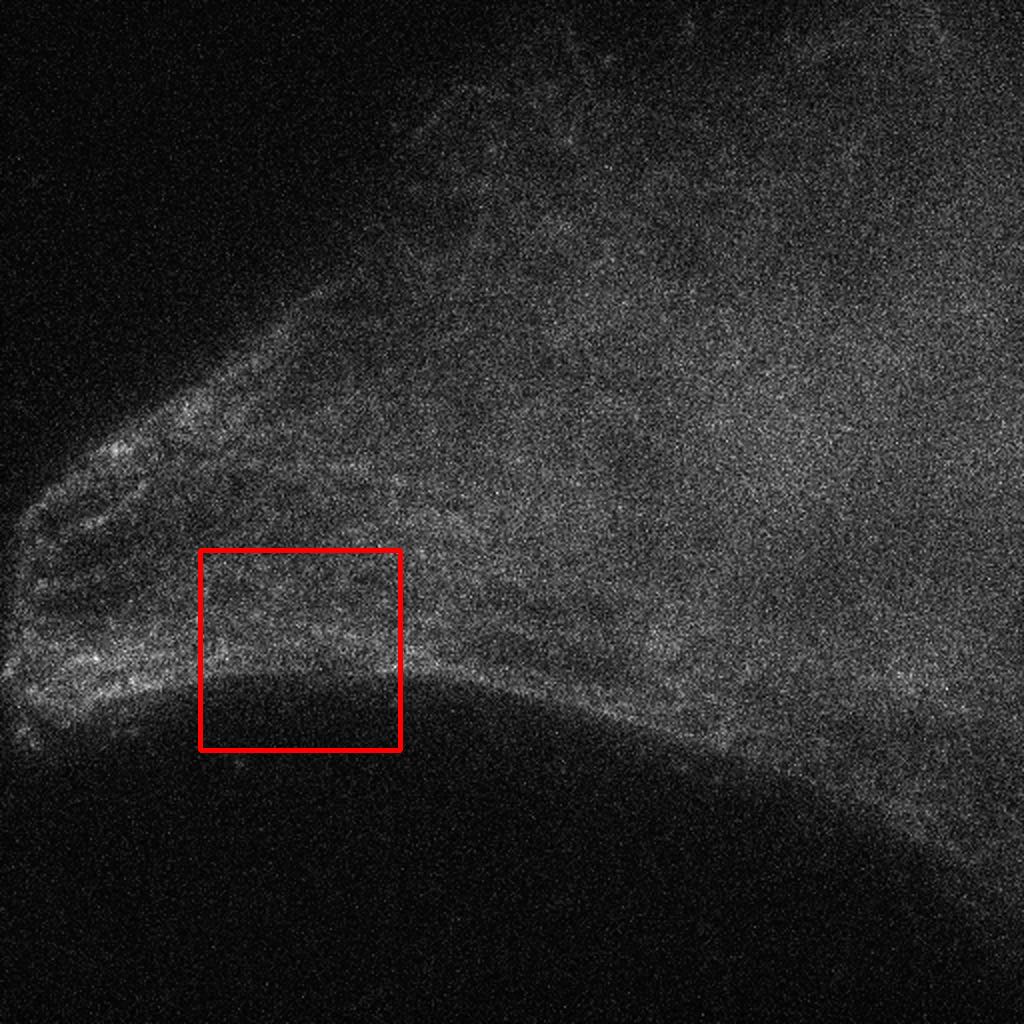}
\end{subfigure}\hspace*{\fill}
\begin{subfigure}{0.19\textwidth}
\caption{Pixel-wise loss \\ (27.09 / 0.81)}
\includegraphics[width=\linewidth]{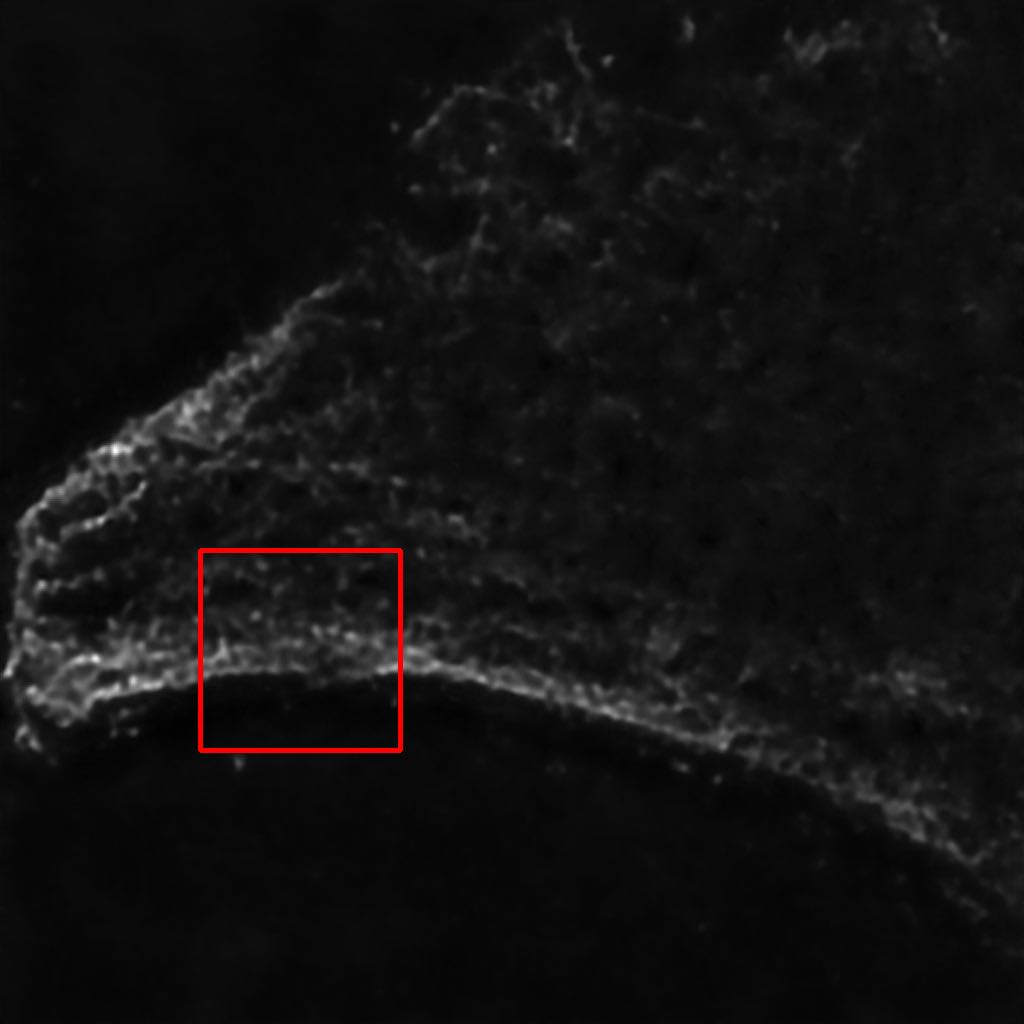}
\end{subfigure}\hspace*{\fill}
\begin{subfigure}{0.19\textwidth}
\caption{Perceptual loss \\ (26.81 / 0.74)}
\includegraphics[width=\linewidth]{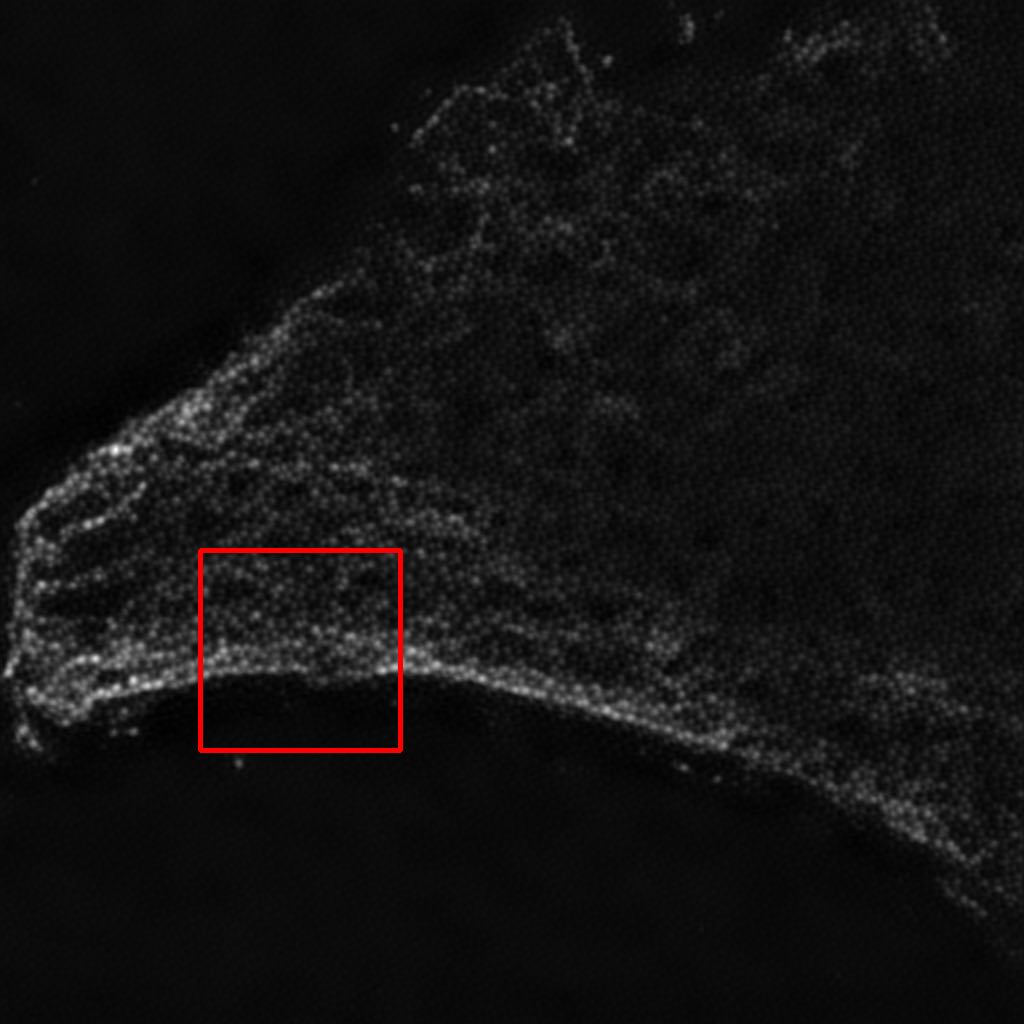}
\end{subfigure}\hspace*{\fill}
\begin{subfigure}{0.19\textwidth}
\caption{GAN loss \\ (28.08 / 0.72)}
\includegraphics[width=\linewidth]{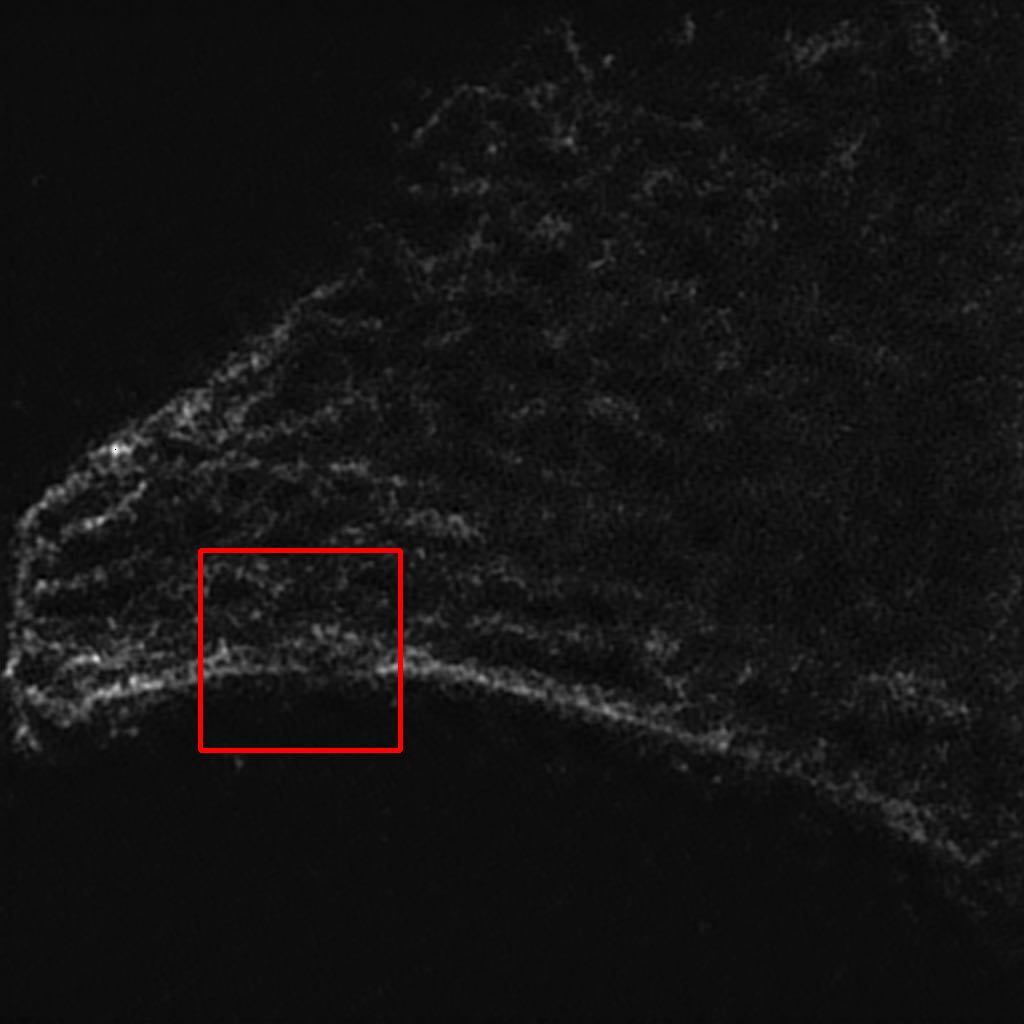}
\end{subfigure}\hspace*{\fill}
\begin{subfigure}{0.19\textwidth}
\caption{SIM \\ \phantom{a}}
\includegraphics[width=\linewidth]{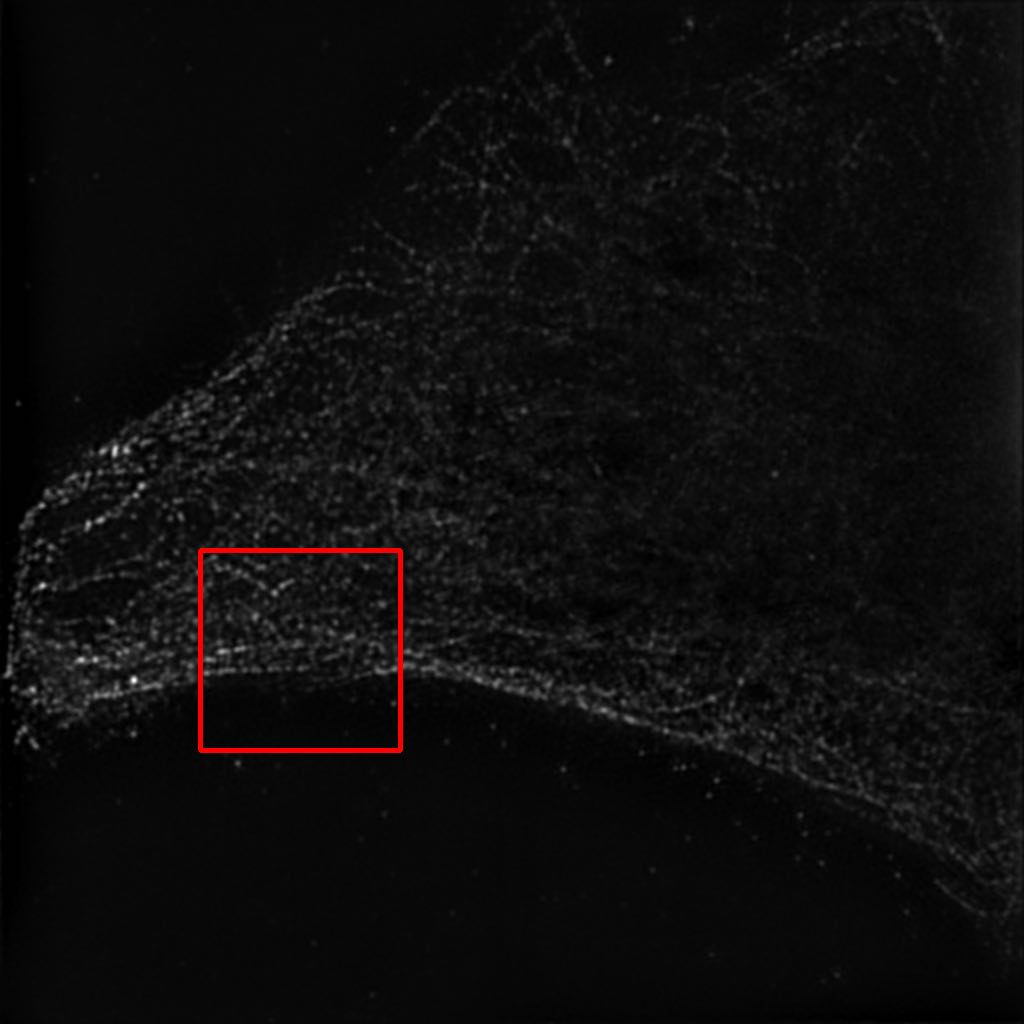}
\end{subfigure}

\medskip

\begin{subfigure}{0.19\textwidth}
\includegraphics[width=\linewidth]{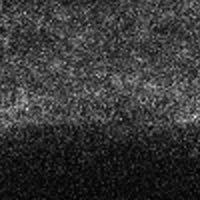}
\end{subfigure}\hspace*{\fill}
\begin{subfigure}{0.19\textwidth}
\includegraphics[width=\linewidth]{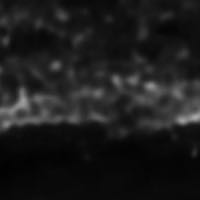}
\end{subfigure}\hspace*{\fill}
\begin{subfigure}{0.19\textwidth}
\includegraphics[width=\linewidth]{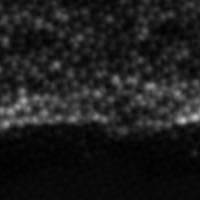}
\end{subfigure}\hspace*{\fill}
\begin{subfigure}{0.19\textwidth}
\includegraphics[width=\linewidth]{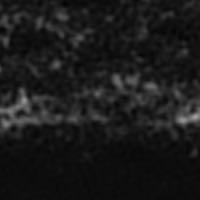}
\end{subfigure}\hspace*{\fill}
\begin{subfigure}{0.19\textwidth}
\includegraphics[width=\linewidth]{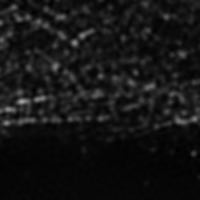}
\end{subfigure}

\caption{Results of our algorithm on an unseen image in the W2S dataset \cite{W2S}, second channel. See legend of Figure \ref{fig:inter_c1}.} \label{fig:inter_c2}
\end{figure}

\begin{figure}[H]
\centering

\begin{subfigure}{0.19\textwidth}
\caption{Bicubic \\ (18.61 / 0.18)}
\includegraphics[width=\linewidth]{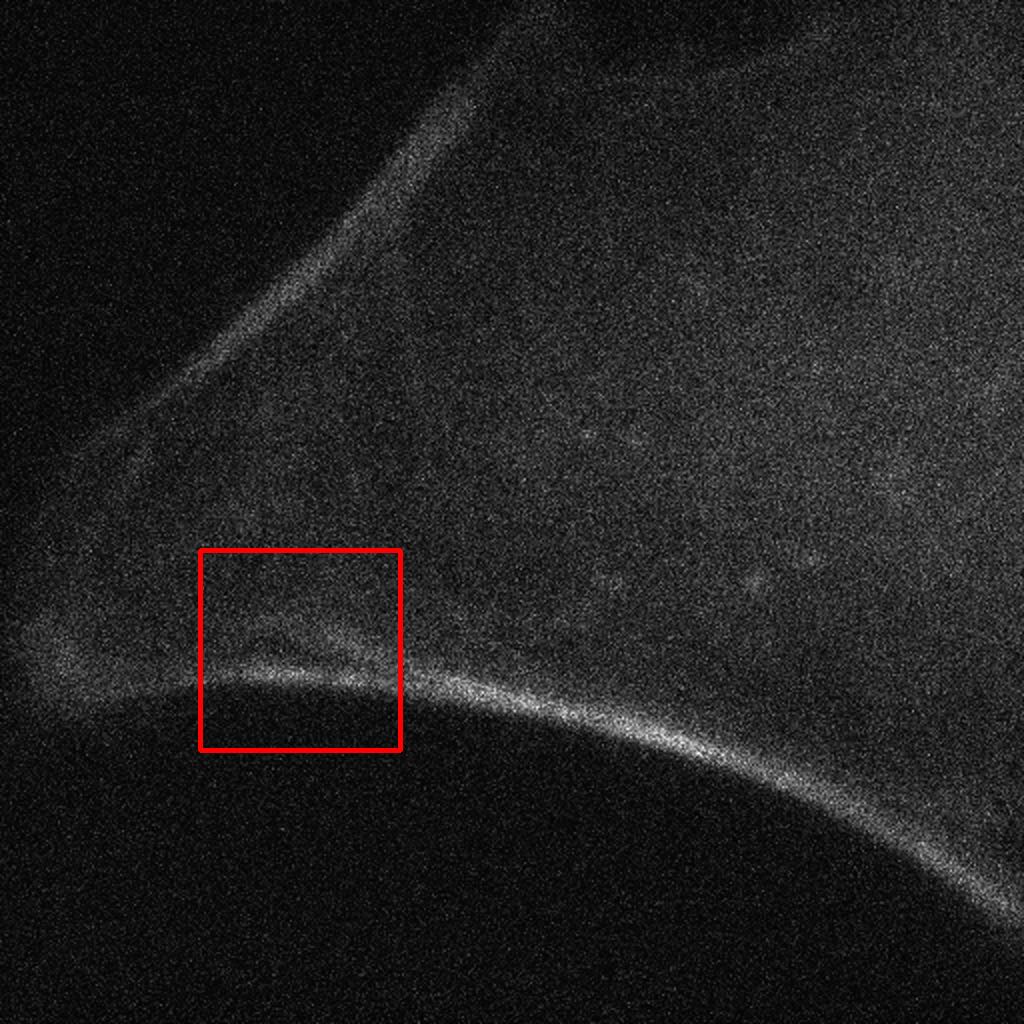}
\end{subfigure}\hspace*{\fill}
\begin{subfigure}{0.19\textwidth}
\caption{Pixel-wise loss \\ (23.90 / 0.49)}
\includegraphics[width=\linewidth]{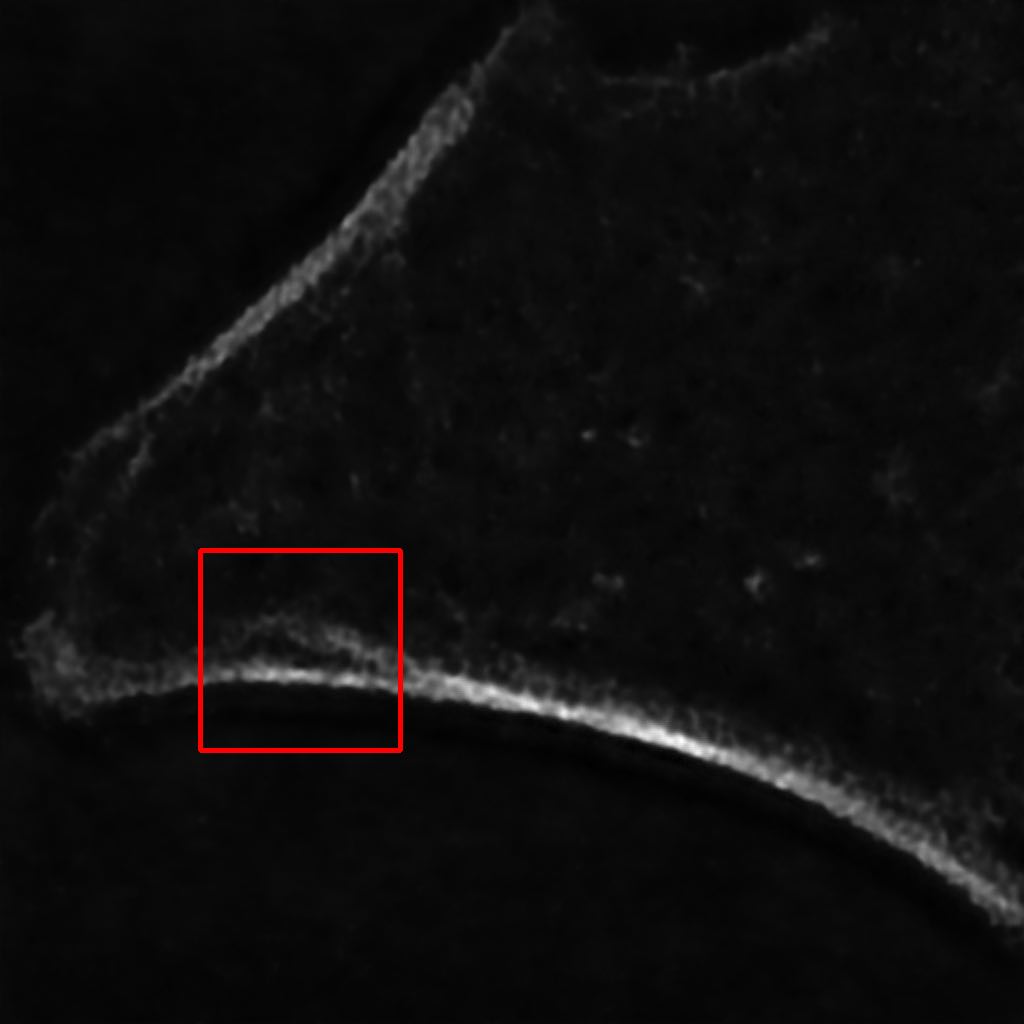}
\end{subfigure}\hspace*{\fill}
\begin{subfigure}{0.19\textwidth}
\caption{Perceptual loss \\ (23.92 / 0.47)}
\includegraphics[width=\linewidth]{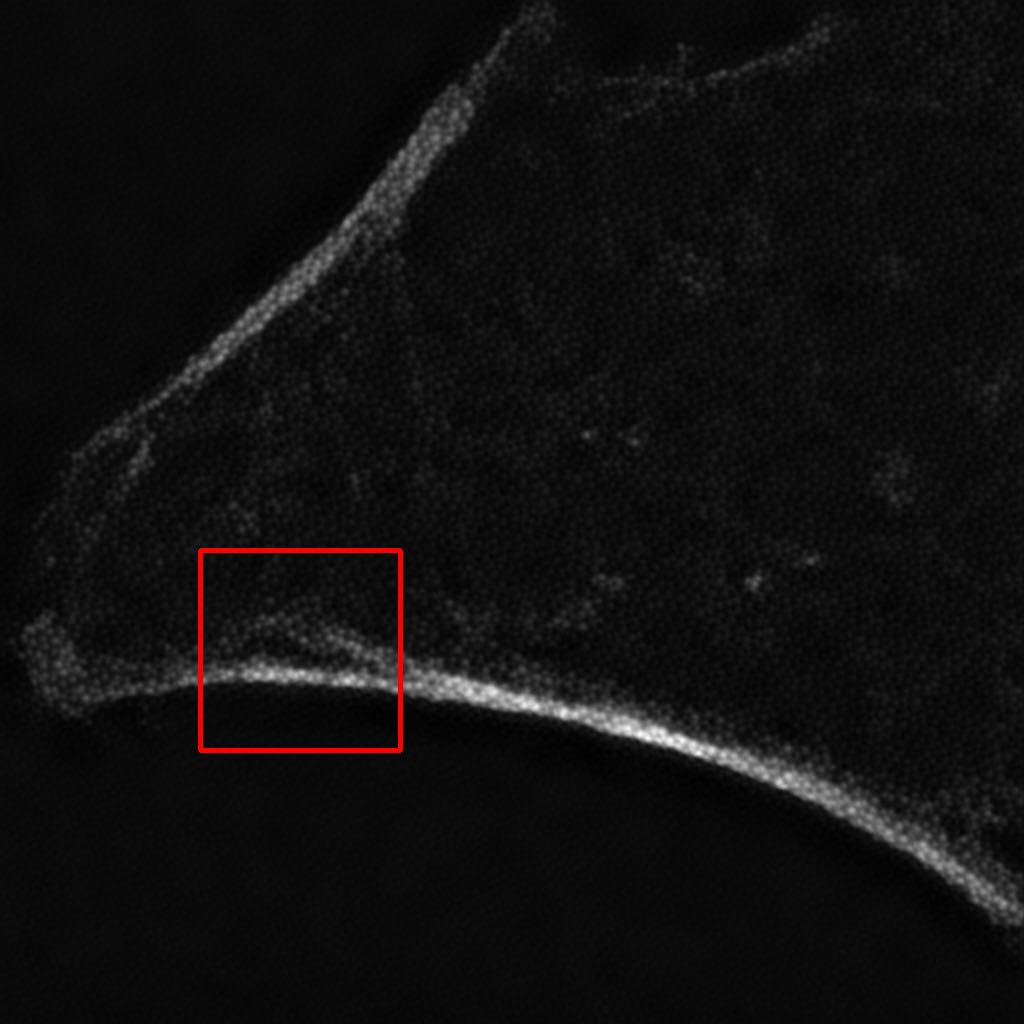}
\end{subfigure}\hspace*{\fill}
\begin{subfigure}{0.19\textwidth}
\caption{GAN loss \\ (22.49 / 0.47)}
\includegraphics[width=\linewidth]{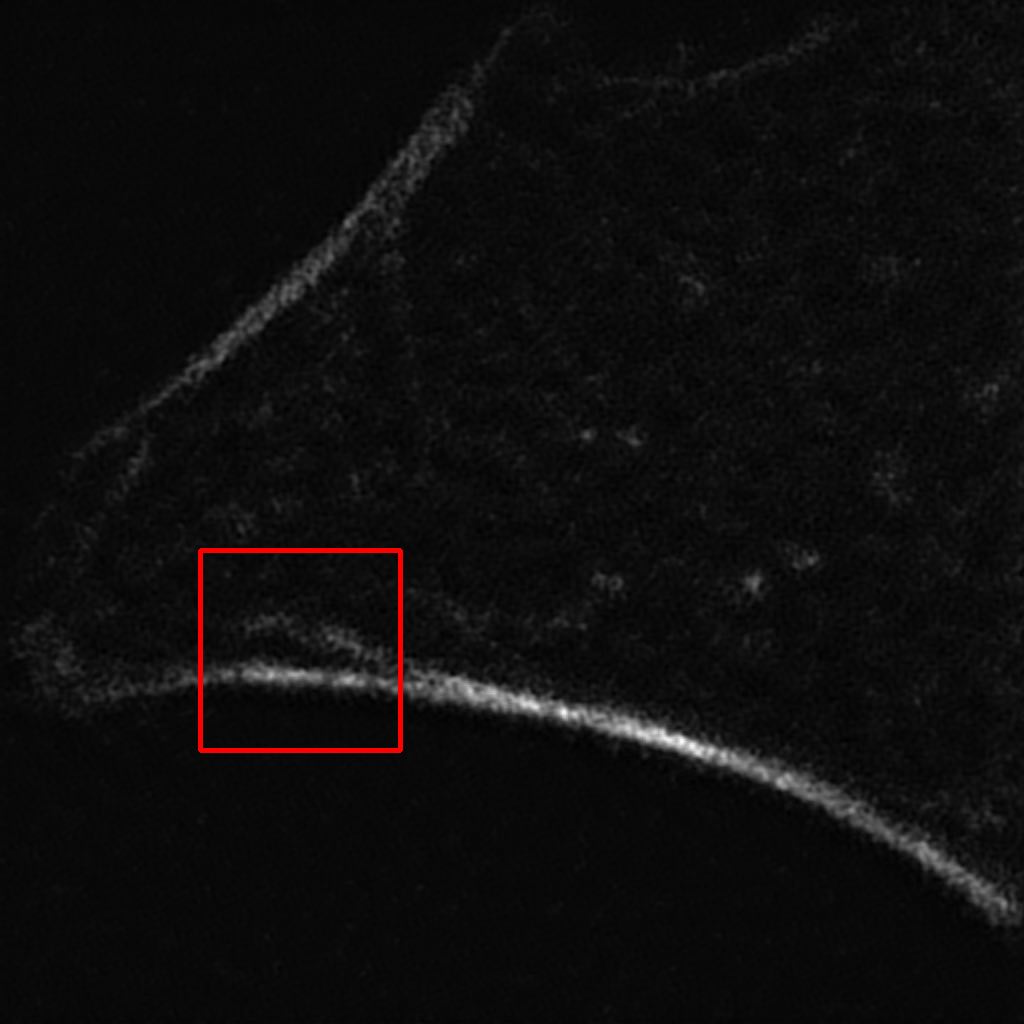}
\end{subfigure}\hspace*{\fill}
\begin{subfigure}{0.19\textwidth}
\caption{SIM \\ \phantom{a}}
\includegraphics[width=\linewidth]{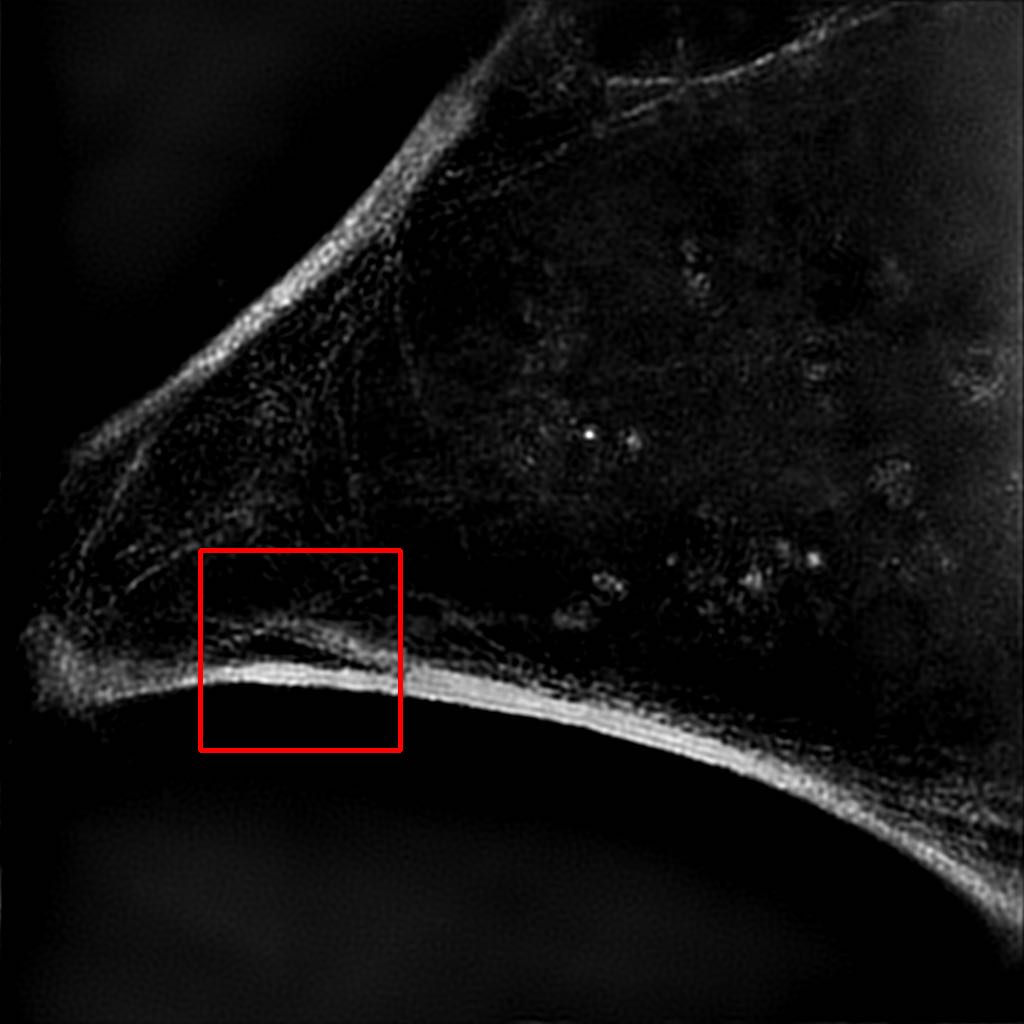}
\end{subfigure}

\medskip

\begin{subfigure}{0.19\textwidth}
\includegraphics[width=\linewidth]{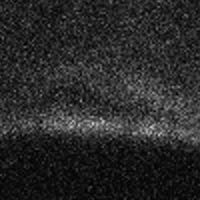}
\end{subfigure}\hspace*{\fill}
\begin{subfigure}{0.19\textwidth}
\includegraphics[width=\linewidth]{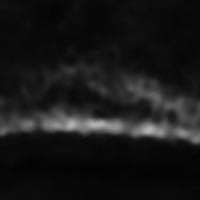}
\end{subfigure}\hspace*{\fill}
\begin{subfigure}{0.19\textwidth}
\includegraphics[width=\linewidth]{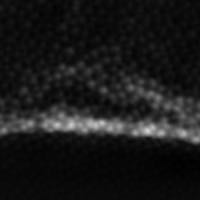}
\end{subfigure}\hspace*{\fill}
\begin{subfigure}{0.19\textwidth}
\includegraphics[width=\linewidth]{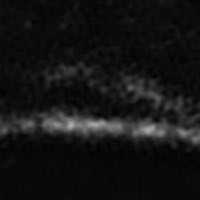}
\end{subfigure}\hspace*{\fill}
\begin{subfigure}{0.19\textwidth}
\includegraphics[width=\linewidth]{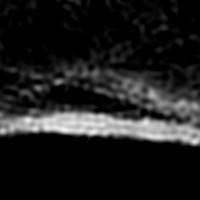}
\end{subfigure}

\caption{Results of our algorithm on an unseen image in the W2S dataset \cite{W2S}, third channel. See legend of Figure \ref{fig:inter_c1}.} \label{fig:inter_c3}
\end{figure}

We can see in Figures \ref{fig:inter_c1},\ref{fig:inter_c2},\ref{fig:inter_c3} that the pre-training phase (panel b) does already apply a significant denoising to the image. However, as was shown in ESRGAN \cite{ESRGAN}, the results obtained with pixel-wise approaches are over-smoothed, as can be seen in the image, that is still blurry. The perceptual approaches with and without the adversarial part (panels d and c) are visually closer to the ground-truth. One can observe that the image without the GAN loss (panel c) is even a bit sharper. This result is also coherent with the observations in ESRGAN \cite{ESRGAN}, that stated that approaches with a strong focus on perceptual loss tend to yield sharper results, but can be accompanied by unpleasant artifacts. This is the reason we prefer the approach with adversarial loss, that is arguably more resistant to such artifacts. We can also observe that neither the PSNR or the SSIM correlate with perceptual quality, as both values are most of the time the highest after the pre-training phase, where the denoising is less effective than after a full training procedure.

\section{Conclusion}

In this project, we show that deep learning algorithms can be efficiently used to enhance the quality of microscopy images. Thanks to recent research on generative adversarial networks and super-resolution, it is now possible to obtain results comparable to state-of-the-art imaging techniques, such as structured illumination microscopy, using a single widefield image acquisition and a pre-trained deep learning model acting as an image processing filter. We use the W2S \cite{W2S} SIM dataset to train a joint denoising and super-resolution model based on ESRGAN \cite{ESRGAN}, that can reproduce results similar as a full SIM experiment pipeline, when applied on a low-resolution widefield microscopy image acquisition. This result is obtained without additional input from the user or further image acquisition metadata. This approach comes with less use time of specialized super-resolution microscopes and better preservation of living cell samples.
\clearpage

{\small
\bibliographystyle{unsrt}
\bibliography{main}
}

\end{document}